\documentclass[journal]{IEEEtran} \newcommand{\figsize}{2.85}
\usepackage{graphicx,psfrag}
\usepackage{url,hyperref}
\usepackage[usenames]{color}
\usepackage[cmex10]{amsmath}
\usepackage{amsfonts,amssymb,latexsym,cite,color,multirow,rotating}
\usepackage{algorithm,algorithmic}
\usepackage{bbm}
\usepackage[normalem]{ulem} 

 \newcommand{\putTable}[3]{\begin{table}[t]
  			    \centering
		            #3
     			    \caption{#2}
     			    \label{tab:#1}
			    \vspace{-4mm}
			  \end{table} }
 \newcommand{\putFrag}[4]{\begin{figure}[t]
                            \centering
                            #4
			    \includegraphics[width=#3in]{figures/#1.eps}
			    \vspace{-2mm}
            		    \caption{#2}
           		    \label{fig:#1}
                          \end{figure} }

 \renewcommand{\hat}{\widehat}
 \renewcommand{\tilde}{\widetilde}
 \newcommand{\defn}{\triangleq}

 \newcommand{\ovec}[1]{\ensuremath{\overline{\boldsymbol{#1}}}}
 \newcommand{\hvec}[1]{\ensuremath{\boldsymbol{\Hat{#1}}}}
    
 \renewcommand{\vec}[1]{\ensuremath{\boldsymbol{#1}}}
 \newcommand{\mat}[1]{\ensuremath{\begin{bmatrix}#1\end{bmatrix}}}
 \newcommand{\smallmat}[1]{\ensuremath{
        \left[\begin{smallmatrix}#1\end{smallmatrix}\right]}}

 \newcommand{\norm}[1]{\ensuremath{\| #1 \|}}
 \newcommand{\mc}[1]{\ensuremath{\mathcal{#1}}}
 \newcommand{\st}{{~\text{s.t.}~}}

 \newcommand{\Real}{{\mathbb{R}}}

 \newcommand{\Nat}{{\mathbb{N}}}

 \newcommand{\tran}{^\textsf{T}}

 \newcommand{\giv}{\,|\,}
 \newcommand{\biggiv}{\,\big|\,}

 \DeclareMathOperator{\E}{E}
 \DeclareMathOperator{\var}{var}

 \DeclareMathOperator{\tr}{tr}

 \DeclareMathOperator*{\argmax}{arg\, max}
 \DeclareMathOperator*{\argmin}{arg\, min}

 \renewcommand{\eqref}[1]{(\ref{eq:#1})}
 
 \newcommand{\Figref}[1]{Figure~\ref{fig:#1}}
 
 \newcommand{\figref}[1]{Fig.~\ref{fig:#1}}
 
 \newcommand{\tabref}[1]{Table~\ref{tab:#1}}
 \newcommand{\secref}[1]{Sec.~\ref{sec:#1}}

 \newcommand{\appref}[1]{Appendix~\ref{app:#1}}

 \newcommand{\ie}{i.e., }

 \newcommand{\textb}[1]{{#1}}

 \newcounter{comment}[section]
 
 \newcounter{texthead}[section]

 \newcommand{\SNR}{\textsf{SNR}}
 \newcommand{\NMSE}{\textsf{NMSE}}
 
 \newcommand{\simp}{\Delta_+^N}
 \newcommand{\rhat}{\hat{r}_n}
 \newcommand{\mur}{\mu^r_n}
 \newcommand{\remove}[1]{_{\backslash #1}}

\newcommand{\naive}{na\"{\i}ve}
 
 \newcommand{\prox}{\text{prox}}

 \newcommand{\X}{\textsf{x}}
 
 \newcommand{\Y}{\textsf{y}}
 \newcommand{\Z}{\textsf{z}}
 
 \renewcommand{\P}{\textsf{p}}
 \newcommand{\R}{\textsf{r}}
 \newcommand{\U}{\textsf{u}}
 \newcommand{\vX}{\textsf{\textbf{\textit{x}}}}
 \newcommand{\mX}{\textsf{\textbf{\textit{X}}}}
 \newcommand{\vY}{\textsf{\textbf{\textit{y}}}}
 
 \newcommand{\vZ}{\textsf{\textbf{\textit{z}}}}
 
 \newcommand{\vR}{\textsf{\textbf{\textit{r}}}}

\begin{document}
\setlength{\arraycolsep}{0.4mm}
 \title{An Empirical-Bayes Approach to Recovering Linearly Constrained Non-Negative Sparse Signals}
	 \author{Jeremy Vila and Philip Schniter\IEEEauthorrefmark{1}
	 \thanks{The authors are with the Department of Electrical and Computer Engineering, The Ohio State University, Columbus, OH.}%
	 \thanks{\IEEEauthorrefmark{1}Please direct all correspondence to Prof. Philip Schniter, Dept. ECE, The Ohio State University, 2015 Neil Ave., Columbus OH 43210, e-mail: schniter@ece.osu.edu, phone 614.247.6488, fax 614.292.7596.}%
	 \thanks{This work has been supported in part by NSF grants IIP-0968910, CCF-1018368, CCF-1218754, and by DARPA/ONR grant N66001-10-1-4090.}
	 \thanks{Portions of this work were presented at the 2013 IEEE Internat. Workshop on Computational Advances in Multi-Sensor Adaptive Processing \cite{Vila:CAMSAP:13}.}
		}
 \date{\today}
 \maketitle

\begin{abstract}
We propose two novel approaches for the recovery of an (approximately) sparse signal from noisy linear measurements in the case that the signal is \textb{a priori} known to be non-negative and obey given linear equality constraints, such as a simplex signal.
This problem arises in, e.g., hyperspectral imaging, portfolio optimization, density estimation, and certain cases of compressive imaging.
Our first approach solves a linearly constrained non-negative version of LASSO using the max-sum version of the generalized approximate message passing (GAMP) algorithm, where we consider both quadratic and absolute loss, and where we propose a novel approach to tuning the LASSO regularization parameter via the expectation maximization (EM) algorithm.
Our second approach is based on the sum-product version of the GAMP algorithm, where we propose the use of a Bernoulli non-negative Gaussian-mixture signal prior and a Laplacian likelihood, and propose an EM-based approach to learning the underlying statistical parameters.
In both approaches, the linear equality constraints are enforced by augmenting GAMP's generalized-linear observation model with noiseless pseudo-measurements.
Extensive numerical experiments demonstrate the state-of-the-art performance of our proposed approaches.
\end{abstract}

\section{Introduction} \label{sec:intro}

We consider the recovery of an (approximately) sparse signal $\vec{x} \in \Real^N$ from the noisy linear measurements 
\begin{equation}
\vec{y} = \vec{Ax} + \vec{w} \in \Real^M ,  \label{eq:y}
\end{equation}
where $\vec{A}$ is a known sensing matrix, $\vec{w}$ is noise, and $M$ may be $\ll N$.
In this paper, we focus on non-negative (NN) signals (i.e., $x_n \geq 0~\forall n$) that obey known linear equality constraints $\vec{Bx} \!=\! \vec{c} \in \Real^P$.
A notable example is \emph{simplex}-constrained signals, i.e., $\vec{x} \in \simp \defn \{ \vec{x} \in \Real^N: x_n \geq 0 \ \forall n, \vec{1}\tran\vec{x} = 1\}$, occurring in hyperspectral image unmixing \cite{Bioucas:JSTAEO:12}, portfolio optimization \cite{Markowitz:Book:91,Brodie:09}, density estimation \cite{Jedynak:NC:2005,Kyrillidis:ICML:13}, and other applications.
We also consider the recovery of NN sparse signals without the linear constraint $\vec{Bx}\!=\!\vec{c}$ \cite{Donoho:PNAS:05b,Bruckstein:TIT:08,Khajehnejad:TSP:11}, which arises in imaging applications \cite{Romberg:SPM:08} and elsewhere \cite{Chen:Chap:09}.

One approach to recovering linearly constrained NN sparse $\vec{x}$ is to solve the $\ell_1$-penalized constrained NN least-squares (LS) problem \eqref{main} 
(see, e.g., \cite{Brodie:09}) for some $\lambda\geq 0$:
\begin{equation}
\label{eq:main}
\hvec{x} = \argmin_{\vec{x} \geq 0} \tfrac{1}{2}\norm{\vec{y} - \vec{Ax}}_2^2 + \lambda \norm{\vec{x}}_1\ \st \ \vec{Bx} = \vec{c}.
\end{equation}
Although this problem is convex \cite{Bertsekas:Book:99}, finding a solution can be computationally challenging in the high-dimensional regime.
Also, while a larger $\lambda$ is known to promote more sparsity in $\hvec{x}$, determining the best choice of $\lambda$ can be difficult in practice.
For example, methods based on cross-validation, the L-curve, or Stein's unbiased risk estimator can be used (see \cite{Giryes:ACHA:11} for discussions of all three), but they require much more computation than solving \eqref{main} for a fixed $\lambda$.
For this reason, \eqref{main} is often considered under the special case $\lambda\!=\!0$ \cite{Slawski:EJS:13}, where it reduces to linearly constrained NN-LS.

For the recovery of $K$-sparse simplex-constrained signals, a special case of the general problem under consideration, the Greedy Selector and Simplex Projector (GSSP) was proposed in \cite{Kyrillidis:ICML:13}. 
GSSP, an instance of projected gradient descent, iterates
\begin{equation}
\label{eq:GSSP}
\hvec{x}^{i+1} = \mc{P}_K \big( \hvec{x}^i - \textsf{step}^i \,\nabla_{\vec{x}} \norm{\vec{y} - \vec{A\hat{x}^i}}_2^2\big),
\end{equation}
where 
$\mc{P}_K(\cdot)$ is \textb{the} Euclidean projection \textb{onto} the $K$-sparse simplex,
$\hvec{x}^i$ is the iteration-$i$ estimate,
$\textsf{step}^i$ is \textb{the} iteration-$i$ step size, 
and 
$\nabla_{\vec{x}}$ is the gradient w.r.t $\vec{x}$. 
For algorithms of this sort, rigorous approximation guarantees can be derived when $\vec{A}$ obeys the restricted isometry property \cite{Garg:ICML:09}. 
Determining the best choice of $K$ can, however, be difficult in practice.


In this paper, we propose two methods for recovering a linearly constrained NN sparse vector $\vec{x}$ from noisy linear observations $\vec{y}$ of the form \eqref{y}, both of which are based on the Generalized Approximate Message Passing (GAMP) algorithm \cite{Rangan:ISIT:11}, an instance of loopy belief propagation that has close connections to primal-dual optimization algorithms \cite{Rangan:ISIT:13,Rangan:ISIT:14}.
When run in ``max-sum'' mode, GAMP can be used to solve optimization problems of the form 
$\hvec{x}=\arg\min_{\vec{x}}\sum_{m=1}^Mh_m([\vec{Ax}]_m)+\sum_{n=1}^Ng_n(x_n)$, where $\hvec{x}$ can be interpreted as the \emph{maximum a posteriori} (MAP) estimate of $\vec{x}$ under the assumed signal prior \eqref{prior} and likelihood \eqref{likelihood}: 
\begin{align}
f(\vec{x}) &\propto \textstyle \prod_{n=1}^N \exp(-g_n(x_n)) \label{eq:prior}\\
f(\vec{y}|\vec{Ax}) &\propto \textstyle \prod_{m=1}^M \exp(-h_m([\vec{Ax}]_m)). \label{eq:likelihood}
\end{align}
When run in ``sum-product'' mode, GAMP returns an approximation of the minimum mean-squared error (MMSE) estimate of $\vec{x}$ under the same assumptions.
In either case, the linear equality constraints $\vec{Bx}\!=\!\vec{c}$ can be enforced through the use of noiseless pseudo-measurements, as described in the sequel.

The first of our proposed approaches solves \eqref{main} using max-sum GAMP while tuning $\lambda$ using a novel expectation-maximization (EM) \cite{Dempster:JRSS:77} procedure.
We henceforth refer to this approach as EM-NNL-GAMP, where NNL is short for ``non-negative LASSO.\footnote{In the absence of the constraint $\vec{Bx}\!=\!\vec{c}$, the optimization problem \eqref{main} can be recognized as a non-negatively constrained version of the LASSO \cite{Tibshirani:JRSSb:96} (also known as basis-pursuit denoising \cite{Chen:JSC:98}).  Similarly, in the special case of $\lambda\!=\!0$, \eqref{main} reduces to non-negative LS \cite{Slawski:EJS:13}.}''
We demonstrate, via extensive numerical experiments, that 
1) the \emph{runtime} of our approach is much faster than the state-of-the-art TFOCS solver \cite{Becker:MPC:11} for a fixed $\lambda$, and that
2) the MSE \emph{performance} of our $\lambda$-tuning procedure is on par with TFOCS under \emph{oracle} tuning. 
We also consider the special case of $\lambda\!=\!0$, yielding ``non-negative least squares GAMP'' (\textb{NNLS-GAMP}), 
whose performance and runtime compare favorably to Matlab's \texttt{lsqlin} routine.
In addition, we consider a variation on \eqref{main} that replaces the quadratic loss $\frac{1}{2}\|\vec{y}-\vec{Ax}\|_2^2$ with the absolute loss $\|\vec{y}-\vec{Ax}\|_1$ for improved robustness to outliers in $\vec{w}$ \cite{Bloomfield:Book:84}, and demonstrate the potential advantages of this technique on a practical dataset.

The second of our proposed approaches aims to solve not an optimization problem like \eqref{main} but rather an inference problem: compute the \emph{MMSE estimate} of a linearly constrained NN sparse vector $\vec{x}$ from noisy linear observations $\vec{y}$.
This is in general a daunting task, since computing the true MMSE estimate requires i) knowing both the true signal prior $f(\vec{x})$ and likelihood $f(\vec{y}|\vec{Ax})$, which are rarely available in practice, and ii) performing optimal inference w.r.t that prior and likelihood, which is rarely possible in practice for computational reasons.

However, when the coefficients in $\vec{x}$ are i.i.d and the observation matrix $\vec{A}$ in \eqref{y} is sufficiently large and random, recent work \cite{Vila:TSP:13} has demonstrated that near-MMSE estimation is indeed possible via the following methodology: place an i.i.d Gaussian-mixture (GM) model with parameters $\vec{q}$ on the coefficients $\{x_n\}$, run sum-product GAMP based on that model, and tune the model parameters $\vec{q}$ using an appropriately designed EM algorithm.
For such $\vec{A}$, the asymptotic optimality of GAMP as an MMSE-inference engine was established in \cite{Rangan:ISIT:11,Javanmard:IMA:13}, and the ability of EM-GAMP to achieve consistent estimates of $\vec{q}$ was established in \cite{Kamilov:NIPS:12}.

In this work, we show that the EM-GM-GAMP approach from \cite{Vila:TSP:13} can be extended to \emph{linearly constrained non-negative} signal models through the use of a non-negative Gaussian-mixture (NNGM) model and noiseless pseudo-measurements, and we detail the derivation and implementation of the resulting algorithm.
Moreover, we demonstrate, via extensive numerical experiments, that EM-NNGM-GAMP's reconstruction MSE is state-of-the-art and that its runtime compares favorably to existing methods.

Both of our proposed approaches can be classified as ``empirical-Bayes'' \cite{Efron:Book:10} in the sense that they combine Bayesian and frequentist approaches: 
GAMP performs (MAP or MMSE) Bayesian inference with respect to a given prior, where the parameters of the prior are treated as deterministic and learned using the EM algorithm, a maximum-likelihood (ML) approach.

\emph{Notation}:
  For matrices, we use boldface capital letters like $\vec{A}$, and we use $\vec{A}\tran$, $\tr(\vec{A})$, and $\norm{\vec{A}}_F$ to denote the transpose, trace, and Frobenius norm, respectively.
  For vectors, we use boldface small letters like $\vec{x}$, and we use $\norm{\vec{x}}_p=(\sum_n |x_n|^p)^{1/p}$ to denote the $\ell_p$ norm, with $x_n=[\vec{x}]_n$ representing the $n^{th}$ element of $\vec{x}$.
  Deterministic quantities are denoted using serif typeface (e.g., $x,\vec{x},\vec{X}$), while random quantities are denoted using san-serif typeface (e.g., $\X,\vX,\mX$).
  For random variable $\X$, we write the pdf as $f_{\X}(x)$, the expectation as $\E\{\X\}$, and the variance as $\var\{\X\}$.
  For a Gaussian random variable $\X$ with mean $m$ and variance $v$, we write the pdf as $\mc{N}(x;m,v)$ 
  and, for the special case of $\mc{N}(x;0,1)$, we abbreviate the pdf as $\varphi(x)$ and write the complimentary cdf as $\Phi_c(x)$.
  Meanwhile, for a Laplacian random variable $\X$ with location $m$ and scale $v$, we write the pdf as $\mc{L}(x;m,v)$.
  For the point mass at $x=0$, we use the Dirac delta distribution $\delta(x)$. 
  Finally,
  we use $\Real$ for the real field 
  and $\int_+g(x) dx$ for the integral of $g(x)$ over $x\in[0,\infty)$.

\section{GAMP overview} \label{sec:GAMP}

As described in \secref{intro}, the generalized approximate message passing (GAMP) algorithm \cite{Rangan:ISIT:11} is an inference algorithm capable of computing either MAP or approximate-MMSE estimates of $\vec{x}\in\Real^N$, where $\vec{x}$ is a realization of random vector $\vX$ with a prior of the form \eqref{prior2}, from generalized-linear observations $\vec{y}\in\Real^M$ that yield a likelihood of the form \eqref{likelihood2},
\begin{align}
f_{\vX}(\vec{x}) &\propto \textstyle \prod_{n=1}^N f_{\X_n}(x_n) \label{eq:prior2}\\
f_{\vY|\vZ}(\vec{y}\giv\vec{Ax}) &\propto \textstyle \prod_{m=1}^M f_{\Y_m|\Z_m}(y_m\giv[\vec{Ax}]_m), \label{eq:likelihood2}
\end{align}
where $\vZ\defn \vec{A}\vX$ represents ``noiseless'' transform outputs.

GAMP generalizes Donoho, Maleki, and Montanari's Approximate Message Passing (AMP) algorithms \cite{Donoho:PNAS:09,Donoho:ITW:10b} from the case of AWGN-corrupted linear observations to the generalized-linear model \eqref{likelihood2}. 
As we shall see, this generalization is useful when enforcing the linear equality constraints $\vec{Bx}\!=\!\vec{c}$ and when formulating non-quadratic variations of \eqref{main}.

GAMP is derived from particular approximations of loopy belief propagation (based on Taylor-series and central-limit-theorem arguments) that yield computationally simple ``first-order'' algorithms bearing strong similarity to primal-dual algorithms \cite{Rangan:ISIT:13,Rangan:ISIT:14}.
Importantly, GAMP admits rigorous analysis in the large-system limit (i.e., $M,N\rightarrow\infty$ for fixed ratio $M/N$) under i.i.d sub-Gaussian $\vec{A}$ \cite{Rangan:ISIT:11,Javanmard:IMA:13}, where its iterations obey a state evolution whose fixed points are optimal whenever they are unique.
Meanwhile, for finite-sized problems and generic $\vec{A}$, max-sum GAMP yields the MAP solution whenever it converges, whereas sum-product GAMP minimizes a certain mean-field variational objective \cite{Rangan:ISIT:13}.
Although performance guarantees for generic finite-dimensional $\vec{A}$ are lacking except in special cases (e.g., \cite{Rangan:ISIT:14}), in-depth empirical studies have demonstrated that (G)AMP performs relatively well for the $\vec{A}$ typically used in compressive sensing applications (see, e.g., \cite{Vila:TSP:13}). 

\tabref{gamp} summarizes the GAMP algorithm.  
\textb{Effectively, GAMP converts the computationally intractable MAP and MMSE high-dimensional vector inference problems to a sequence of scalar inference problems.
In the end,} its complexity is dominated by four\footnote{Two matrix multiplies per iteration, those in (R1) and (R9), can be eliminated using the ``scalar variance'' modification of GAMP, with vanishing degradation in the large-system limit \cite{Rangan:ISIT:11}.} 
matrix-vector multiplies per iteration: steps (R1), (R2), (R9), (R10). 
Furthermore, GAMP can take advantage of fast implementations of the matrix-vector multiplies (e.g., FFT) when they exist.
For max-sum GAMP, scalar inference is accomplished by lines (R3) and (R11), which involve the proximal operator
\begin{equation} 
\label{eq:prox}
\prox_g(\hat{v};\mu^v) \defn \argmin_{x \in \Real} g(x) + \frac{1}{2\mu^v}|x-\hat{v}|^2
\end{equation}
for generic scalar function $g(\cdot)$, as well as lines (R4) and (R12), which involve the derivative of the $\prox$ operator \eqref{prox} with respect to its first argument.  
Meanwhile, for sum-product GAMP, scalar inference is accomplished by lines (R5) and (R6), which compute the mean and variance of GAMP's iteration-$t$ approximation to the marginal posterior on $\Z_m$,
\begin{align}
  f_{\Z_m|\P_m}\!(z\giv\hat{p}_m(t); \mu^p_m(t))
  &\propto f_{\Y_m|\Z_m}(y_m|z) \mc{N}(z;\hat{p}_m(t),\mu^p_m(t)),
  	\label{eq:gamppostz} 
\end{align}
and by lines (R13) and (R14), which compute the mean and variance of the GAMP-approximate marginal posterior on $\X_n$,
\begin{align}
  f_{\X_n|\R_n}\!(x\giv\hat{r}_n(t); \mu^r_n(t)) 
  &\propto f_{\X_n}(x) \mc{N}(x; \hat{r}_n(t), \mu^r_n(t)).
	\label{eq:gamppost}
\end{align}

\textb{
We now provide background on GAMP that helps to explain \eqref{gamppostz}-\eqref{gamppost} and \tabref{gamp}.
First and foremost, GAMP can be interpreted as an iterative thresholding algorithm, in the spirit of, e.g., \cite{Chambolle:TIP:98,Daubechies:CPAM:04}.
In particular, when the GAMP-assumed distributions are matched to the true ones, the variable $\hat{r}_n(t)$ produced in (R10) is \emph{an approximately AWGN-corrupted version of the true coefficient $x_n$} (i.e., $\hat{\R}_n(t)=\X_n+\tilde{\R}_n(t)$ with $\tilde{\R}_n(t)\sim\mc{N}(0,\mu_n^r(t))$ independent of $\X_n$) where $\mu_n^r(t)$ is computed in (R9) and the approximation becomes exact in the large-system limit with i.i.d sub-Gaussian $\vec{A}$ \cite{Rangan:ISIT:11,Javanmard:IMA:13}.
Note that, under this AWGN corruption model, the pdf of $\X_n$ given $\hat{\R}_n(t)$ takes the form in \eqref{gamppost}.
Thus, in sum-product mode, GAMP sets $\hat{x}_n(t\!+\!1)$ at the scalar MMSE estimate of $\X_n$ given $\hat{r}_n(t)$, as computed via the conditional mean in (R13), and it sets $\mu^x_n(t\!+\!1)$ as the corresponding MMSE, as computed via the conditional variance in (R14).
Meanwhile, in max-sum mode, GAMP sets $\hat{x}_n(t\!+\!1)$ at the scalar MAP estimate of $\X_n$ given $\hat{r}_n(t)$, as computed by the prox step in (R11), and it sets $\mu^x_n(t\!+\!1)$ in accordance with the sensitivity of this proximal thresholding, as computed in (R12). 
This explains \eqref{gamppost} and lines (R9)-(R14) in \tabref{gamp}.}

\textb{
We now provide a similar explanation for \eqref{gamppostz} and lines (R1)-(R6) in \tabref{gamp}.
When the GAMP distributions are matched to the true ones, $\hat{p}_m(t)$ produced in (R2) is \emph{an approximately AWGN-corrupted version of the true transform output $z_m$} (i.e., $\hat{\P}_m(t)=\Z_m+\tilde{\P}_m(t)$ with $\tilde{\P}_m(t)\sim\mc{N}(0,\mu_m^p(t))$ independent of $\hat{\P}_m(t)$) where $\mu_m^p(t)$ is computed in (R1) and the approximation becomes exact in the large-system limit with i.i.d sub-Gaussian $\vec{A}$ \cite{Rangan:ISIT:11,Javanmard:IMA:13}.
Under this model, the pdf of $\Z_m$ given $\hat{\P}_m(t)$ and $\Y_m$ takes the form in \eqref{gamppostz}.
Thus, in sum-product mode, GAMP sets $\hat{z}_m(t)$ at the scalar MMSE estimate of $\Z_m$ given $\hat{p}_m(t)$ and $y_m$, as computed via the conditional mean in (R5), and it sets $\mu^z_m(t)$ as the corresponding MMSE, as computed via the conditional variance in (R6).
Meanwhile, in max-sum mode, GAMP sets $\hat{z}_m(t)$ at the scalar MAP estimate of $\Z_m$ given $\hat{p}_m(t)$ and $y_m$, as computed by the prox operation in (R3), and it sets $\mu^z_m(t)$ in accordance with the sensitivity of this prox operation, as computed in (R4).}

\textb{
Indeed, what sets GAMP (and its simpler incarnation AMP) apart from other iterative thresholding algorithms is that the thresholder inputs $\hat{r}_n(t)$ and $\hat{p}_m(t)$ are (approximately) AWGN corrupted observations of $\X_n$ and $\Z_m$, respectively, ensuring that the scalar thresholding steps (R3)-(R6) and (R11)-(R14) are well justified from the MAP or MMSE perspectives.
Moreover, it is the ``Onsager'' correction ``$-\mu^p_m(t)\hat{s}_m(t\!-\!1)$'' in (R2) that ensures the AWGN nature of the corruptions; without it, AMP reduces to classical iterative thresholding \cite{Donoho:PNAS:09}, which performs much worse \cite{Maleki:JSTSP:10}.
Computing the Onsager correction involves (R7)-(R8). 
To our knowledge, the simplest interpretation of the variables $\hat{s}_m(t)$ and $\mu^s_m(t)$ computed in (R7)-(R8) comes from primal-dual optimization theory, as established in \cite{Rangan:ISIT:14}: whereas $\hat{x}_n(t)$ are estimates of the primal variables, $\hat{s}_m(t)$ are estimates of the dual variables; and whereas $\mu^x_n(t)$ relates to the primal sensitivity at the point $\hat{x}_n(t)$, $\mu^s_m(t)$ relates to the dual sensitivity at $\hat{s}_m(t)$.
}

\putTable{gamp}{The GAMP Algorithm from \cite{Rangan:ISIT:11} with max iterations $T_{\max}$ and stopping tolerance $\epsilon_{\textsf{gamp}}$.}
{\footnotesize
\begin{equation*}
\begin{array}{|lrcl@{}r|}\hline
  \multicolumn{5}{|l|}{\textsf{inputs:~~}\forall m,n:
  	f_{\X_n}, 
	f_{\Y_m|\Z_m}, 
	A_{mn}, T_{\max}, \epsilon_{\textsf{gamp}}>0,
	\textsf{MaxSum}\in\{0,1\}\!
	}\\[1mm]
  \multicolumn{2}{|l}{\textsf{definitions:}}&&&\\[-1mm]
  &f_{\Z_m|\P_m}(z\giv \hat{p};\mu^p) \!
   &\defn& \frac{f_{\Y_m|\Z_m}(y_m|z) \,\mc{N}(z;\hat{p},\mu^p)}
	{\int_{z} f_{\Y_m|\Z_m}(y_m|z) \,\mc{N}(z;\hat{p},\mu^p)} &\text{\scriptsize(D1)}\\
  &f_{\X_n|\R_n}(x\giv \hat{r};\mu^r) \!
   &\defn& \frac{f_{\X_n}\!(x) \,\mc{N}(x;\hat{r},\mu^r)}
        {\int_{x}f_{\X_n}\!(x) \,\mc{N}(x;\hat{r},\mu^r)}&\text{\scriptsize(D2)}\\
  \multicolumn{2}{|l}{\textsf{initialize:}}&&&\\
  &\forall n: 
   \hat{x}_{n}(1) &=& \int_{x} x\, f_{\X_n}\!(x) & \text{\scriptsize(I1)}\\
  &\forall n:
   \mu^x_n(1) &=& \int_{x} |x-\hat{x}_n(1)|^2  f_{\X_n}\!(x) & \text{\scriptsize(I2)}\\
  &\forall m: 
   \hat{s}_{m}(0) &=& 0 & \text{\scriptsize(I3)}\\
  \multicolumn{2}{|l}{\textsf{for $t=1:T_{\max}$,}}&&&\\
  &\forall m:
   \mu^p_m(t)
   &=& \textstyle \sum_{n=1}^{N} |\!A_{mn}|^2 \mu^x_{n}(t) & \text{\scriptsize(R1)}\\
  &\forall m:
   \hat{p}_m(t)
   &=& \sum_{n=1}^{N} \!A_{mn} \hat{x}_{n}(t) - \mu^p_m(t) \,\hat{s}_m(t-1)& \text{\scriptsize(R2)}\\
   &\multicolumn{3}{l}{\hspace{3.5mm}
      \textsf{if MaxSum then}} & \\
  &\forall m:
   \hat{z}_m(t)
   &=& \prox_{- \ln f_{\Y_m|\Z_m}}\!(\hat{p}_m(t);\mu^p_m(t)) & \text{\scriptsize(R3)}\\
  &\forall m:
   \mu^z_m(t)
  &=&\mu^p_m(t) \,\prox^\prime_{- \ln f_{\Y_m|\Z_m}}\!(\hat{p}_m(t);\mu^p_m(t)) & \text{\scriptsize(R4)}\\
  &\multicolumn{3}{l}{\hspace{3.5mm}
      \textsf{else}} & \\
  &\forall m:
   \hat{z}_m(t)
   &=& \E\{\Z_m|\P_m \!=\! \hat{p}_m(t);\mu^p_m(t)\} & \text{\scriptsize(R5)}\\
  &\forall m:
   \mu^z_m(t)
   &=& \var\{\Z_m|\P_m \!=\! \hat{p}_m(t);\mu^p_m(t)\} & \text{\scriptsize(R6)}\\
   &\multicolumn{3}{l}{\hspace{3.5mm}
      \textsf{end if}} & \\
  &\forall m:
   \mu^s_m(t)
   &=& \big(1-\mu^z_m(t)/\mu^p_m(t)\big)/\mu^p_m(t) & \text{\scriptsize(R7)}\\
  &\forall m:
   \hat{s}_m(t)
   &=& \big(\hat{z}_m(t)-\hat{p}_m(t)\big)/\mu^p_m(t) & \text{\scriptsize(R8)}\\
  &\forall n:
   \mu^r_n(t)
   &=& \textstyle \big(\sum_{m=1}^{M} |\!A_{mn}|^2 \mu^s_m(t) 
	\big)^{-1} & \text{\scriptsize(R9)}\\
  &\forall n:
   \hat{r}_n(t)
   &=& \textstyle \hat{x}_n(t)+ \mu^r_n(t) \sum_{m=1}^{M} \!A_{mn}^*
	\hat{s}_{m}(t)  & \text{\scriptsize(R10)}\\
   &\multicolumn{3}{l}{\hspace{3.5mm}
      \textsf{if MaxSum then}} & \\
   &\forall n:
   \hat{x}_n(t\!+\!1)
   &=& \prox_{- \ln f_{\X_n}}\!(\hat{r}_n(t);\mu^r_n(t)) & \text{\scriptsize(R11)}\\
   &\forall n:
   \mu^x_n(t\!+\!1)
   &=&\mu^r_n(t)\,\prox^\prime_{- \ln f_{\X_n}}\!(\hat{r}_n(t);\mu^r_n(t)) & \text{\scriptsize(R12)}\\
  &\multicolumn{3}{l}{\hspace{3.5mm}
      \textsf{else}} & \\
  &\forall n:
   \hat{x}_{n}(t\!+\!1)
   &=& \E\{\X_n|\R_n\!=\!\hat{r}_n(t);\mu^r_n(t)\} & \text{\scriptsize(R13)}\\
  &\forall n:
   \mu^x_n(t\!+\!1)
   &=& \var\{\X_n|\R_n\!=\!\hat{r}_n(t);\mu^r_n(t)\} & \text{\scriptsize(R14)}\\
  &\multicolumn{3}{l}{\hspace{3.5mm}
      \textsf{end if}} & \\
  &\multicolumn{3}{l}{\hspace{3.5mm}
      \textsf{if} \sum_{n=1}^N|\hat{x}_n(t\!+\!1)-\hat{x}_n(t)|^2 
  	< \epsilon_{\textsf{gamp}} \sum_{n=1}^N|\hat{x}_n(t)|^2,
      \textsf{break}} & \text{\scriptsize(R15)}\\
  \multicolumn{2}{|l}{\textsf{end}}&&&\\[1mm]
  \multicolumn{4}{|l}{\textsf{outputs:~~}
        \forall m,n:
	\hat{z}_m(t),\mu^z_m(t),
	\hat{r}_n(t),\mu^r_n(t), 
	\hat{x}_n(t\!+\!1),\mu^x_n(t\!+\!1) 
	}&\\[1mm]
  \hline
\end{array}
\end{equation*}
\vspace{-2 mm}
}

\section{Observation Models}	\label{sec:sys_model}

To enforce the linear equality constraint $\vec{Bx} \!=\! \vec{c} \in\Real^P$ using GAMP, we extend the observation model \eqref{y} to 
\begin{equation}
\mat{\vec{y}\\\vec{c}}
= \mat{\vec{A} \\ \vec{B}} \vec{x} + \mat{\vec{w} \\ \vec{0}} 
\label{eq:model}
\end{equation}
and exploit the fact that GAMP supports a likelihood function that varies with the measurement index $m$.
Defining 
$\ovec{y}\defn \mat{\vec{y}\tran &\vec{c}\tran}\tran$,
$\ovec{A}\defn \mat{\vec{A}\tran&\vec{B}\tran}\tran$, and
$\ovec{z}\defn \ovec{A}\vec{x}$, 
the likelihood associated with the augmented model \eqref{model} can be written as
\begin{align}
  \label{eq:pY_Z}
  &\\[-6mm]
  f_{\overline \Y_{m}|\overline \Z_{m}}(\overline y_{m}|\overline z_{m})
  \!=\!\begin{cases}
    f_{\Y|\Z}(\overline y_m|\overline z_m)
    & m=1,\dots,M\\
    \delta(\overline{y}_m\!-\overline z_{m}) 
    & m=M\!+\!1,\dots, M\!+\!P,
  \end{cases}
  \nonumber
\end{align}
where $f_{\Y|\Z}$ is the likelihood of the first $M$ measurements \eqref{y}.

Note that, for either max-sum or sum-product GAMP, the quantities in (R3)-(R6) of \tabref{gamp} then become
\begin{align}
 \hat{z}_m(t) &= c_{m-M} &m=M\!+1,\dots,M\!+\!P \label{eq:augzhat}\\
 \mu^z_m(t)  &= 0 &m = M\!+1,\dots,M\!+\!P \label{eq:augzvar},
\end{align}
\textb{where $c_{m-M}$ are elements of $\vec{c}$.}


\subsection{Additive white Gaussian noise} \label{sec:AWGNout}

When the noise $\vec{w}$ is modeled as \textb{additive white Gaussian noise} (AWGN) with variance $\psi$, the likelihood $f_{\Y|\Z}$ in \eqref{pY_Z} takes the form
\begin{equation}
\label{eq:AWGN}
f_{\Y|\Z}(y|z) = \mc{N}(y;z,\psi).
\end{equation}
In this case, for either max-sum or sum-product GAMP, the quantities in (R3)-(R6) of \tabref{gamp} become \cite{Rangan:ISIT:11} (omitting the $t$ index for brevity)
\begin{align}
\label{eq:zhatAWGN} 
 \hat{z}_m &= \hat{p}_m + \tfrac{\mu^p_m}{\mu^p_m+\psi}(y_m - \hat{p}_m) &m = 1,\dots,M\phantom{.}\\
\label{eq:zvarAWGN}
 \mu^z_m &= \frac{\mu^p_m\psi}{\mu^p_m + \psi} & m = 1,\dots,M. 
\end{align}

\subsection{Additive white Laplacian noise} \label{sec:LAPout}

The additive white Laplacian noise (AWLN) observation model is an alternative to the AWGN model that is more robust to outliers \cite{Bloomfield:Book:84}.
\textb{Here, the noise $\vec{w}$ is modeled as AWLN with} rate parameter $\psi>0$, and the corresponding likelihood $f_{\Y|\Z}$ in \eqref{pY_Z} takes the form
\begin{equation}
\label{eq:LAP}
f_{\Y|\Z}(y|z) = \mc{L}(y;z,\psi) \defn \tfrac{\psi}{2}\exp(-\psi |y -z|),
\end{equation}
and so, for the max-sum case, (R3) in \tabref{gamp} becomes
\begin{equation}
\label{eq:softthresh}
\hat{z}_m = \argmin_{z_m\in\Real} |z_m - y_m| + \frac{(z_m-\hat{p}_m)^2}{2 \textb{\mu^p_m}\psi} .
\end{equation}
The solution to \eqref{softthresh} can be recognized as a $y_m$-shifted version of ``soft-thresholding'' function, and so the max-sum quantities in (R3) and (R4) of \tabref{gamp} become, using $\tilde{p}_m \defn\hat{p}_m - y_m$,  
\begin{align}
\label{eq:softthreshout}
\hat{z}_m &=
\begin{cases}
\hat{p}_m - \psi\mu^p_m &\tilde{p}_m \geq \psi\mu^p_m \\
\hat{p}_m + \psi\mu^p_m &\tilde{p}_m \leq -\psi\mu^p_m \\
y_m &\text{else}
\end{cases}
& m = 1,\dots,M, \\
\label{eq:laplacemuz}
\mu^z_m &= 
\begin{cases}
0 & |\tilde{p}_m| \leq \psi\mu^p_m \\
\mu^p_m &\text{else}
\end{cases} & m = 1,\dots,M.
\end{align}
Meanwhile, as shown in \appref{MMSE_Lap}, the sum-product GAMP quantities (R5) and (R6) (i.e., the mean and variance of the GAMP approximated $\Z_m$ posterior \eqref{gamppostz}) become

\small
\begin{align}
\label{eq:MMSELapmean} 
\hat{z}_m &= y_m \!+ \frac{\underline{C}_m}{C_m}\left(\underline{p}_m \!\!-\! \sqrt{\mu^p_m} h(\underline{\kappa}_m) \right)
\!+\! \frac{\overline{C}_m}{C_m} \left(\overline{p}_m \!+\! \sqrt{\mu^p_m} h(\overline{\kappa}_m)\right)\!\! \\
\label{eq:MMSELapvar}
\mu^z_m &= \frac{\underline{C}_m}{C_m} \left( \mu^p_m g(\underline{\kappa}_m) \!+\! \left( \underline{p}_m \!-\! \sqrt{\mu^p_m} h(\underline{\kappa}_m)\right)^2 \right) \\
&\quad+\frac{\overline{C}_m}{C_m} \left( \mu^p_m g(\overline{\kappa}_m) \!+\! \left( \overline{p}_m \!+\! \sqrt{\mu^p_m} h(\overline{\kappa}_m)\right)^2 \right)\! \!-\!(y_m \!-\! \hat{z}_m)^2, \nonumber
\end{align}
\normalsize
where $\underline{p}_m \defn \tilde{p}_m + \psi \mu^p_m$, $\overline{p}_m \defn \tilde{p}_m - \psi \mu^p_m$, 
\begin{align}
\label{eq:CL}
\underline{C}_m &\defn \tfrac{\psi}{2}\exp\left(\psi+ \tfrac{1}{2}\psi^2\mu^p_m \right) \Phi_c(\underline{\kappa}_m)  \\
\label{eq:CU}
\overline{C}_m &\defn \tfrac{\psi}{2}\exp\left(-\psi + \tfrac{1}{2}\psi^2\mu^p_m \right) \Phi_c(\overline{\kappa}_m),
\end{align}
$C_m \defn \underline{C}_m + \overline{C}_m$, $\underline{\kappa}_m \defn \underline{p}_m/\sqrt{\mu^p_m}$, $\overline{\kappa}_m \defn -\overline{p}_m/\sqrt{\mu^p_m}$ and 
\begin{align}
\label{eq:h}
h(a) & \defn \frac{\varphi(a)}{\Phi_c(a)}\\
\label{eq:g}
g(a) & \defn 1- h(a)\big(h(a) - a\big).
\end{align}

\section{Non-Negative GAMP} 										\label{sec:NN_GAMP}


\subsection{NN Least Squares GAMP} 										\label{sec:US_GAMP}

We first detail the \textb{NNLS-GAMP} algorithm, which uses max-sum GAMP to solve the $\lambda = 0$ case of \eqref{main}. 
Noting that the $\vec{x}\geq \vec{0}$ constraint in \eqref{main} can be thought of as adding an infinite penalty to the quadratic term when any $x_n < 0$ and no additional penalty otherwise, we model the elements of $\vec{x}$ as i.i.d random variables with the (improper) NN prior pdf
\begin{equation}
f_\X(x) = 
\begin{cases}
1 &x \geq 0 \\ 0 &x < 0
\end{cases},
\label{eq:unifpri}
\end{equation}
and we assume the augmented model \eqref{pY_Z} with AWGN likelihood \eqref{AWGN} (of variance $\psi=1$), in which case max-sum GAMP performs the unconstrained optimization
\begin{equation}
\label{eq:USGAMPcost}
\argmin_{\vec{x}}
-\!\sum_{n=1}^N \!\ln \mathbbm{1}_{x_n \geq 0} - \ln \mathbbm{1}_{\vec{B}\vec{x} = \vec{c}} + \frac{1}{2}\norm{\vec{y} - \vec{Ax}}_2^2,
\end{equation}
where $\mathbbm{1}_A\in\{0,1\}$ is the indicator function of the event $A$.  
Hence, \eqref{USGAMPcost} is equivalent to the constrained optimization \eqref{main} when $\lambda = 0$.  

Under the i.i.d NN uniform prior \eqref{unifpri}, it is readily shown that the max-sum GAMP steps (R11) and (R12) become
\begin{align}
\hat{x}_n &=
\begin{cases}
 0 & \hat{r}_n \leq 0 \\
\hat{r}_n &\hat{r}_n > 0 
\end{cases}, \\
\mu^x_n &= 
\begin{cases}
0 &\hat{r}_n \leq 0 \\
\mu^r_n &\hat{r}_n > 0
\end{cases}.
\end{align}

\subsection{NN LASSO GAMP} 					\label{sec:NNLGAMP}


Next we detail the \textb{NNL-GAMP} algorithm, which uses max-sum GAMP to solve the $\lambda>0$ case of \eqref{main}.
For this, we again employ the augmented model \eqref{pY_Z} and AWGN likelihood \eqref{AWGN} (\textb{with variance $\psi$}), but \textb{we} now use i.i.d exponential $\X_n$, i.e.,
\textb{
\begin{equation}
\label{eq:pX_NNL}
f_\X(x) = 
\begin{cases}
\chi\exp(-\chi x) &x\geq 0 \\
0 &\text{else}
\end{cases}
\end{equation}
for $\chi>0$.
With these priors and the augmented observation model \eqref{model}, NNL-GAMP solves the optimization problem
\begin{equation}
\label{eq:NNLGAMPcost}
\hvec{x} = \argmin_{\vec{x} \geq 0} \tfrac{1}{2\psi}\norm{\vec{y} - \vec{Ax}}_2^2 + \chi \norm{\vec{x}}_1\ \st \ \vec{Bx} = \vec{c},
\end{equation}
which reduces to \eqref{main} under $\lambda = \chi\psi$.}

\textb{
It is then straightforward to show that the max-sum lines (R11) and (R12) in \tabref{gamp} reduce to 
\begin{align}
\label{eq:expxhat}
\hat{x}_n &=
\begin{cases}
\hat{r}_n - \chi\mu^r_n &\hat{r}_n \geq \chi\mu^r_n \\
0 &\text{else}
\end{cases} \\
\label{eq:expmux}
\mu^x_n &= 
\begin{cases}
\mu^r_n & \hat{r}_n \geq \chi\mu^r_n \\
0 & \text{else} \\
\end{cases}.
\end{align}
}

\subsection{NN Gaussian Mixture GAMP} 				\label{sec:NNGMGAMP}

Finally, we detail the NNGM-GAMP algorithm, which employs sum-product GAMP under the i.i.d Bernoulli non-negative Gaussian mixture (NNGM) prior pdf for $\vX$, i.e.,
\begin{equation}
f_\X(x) = (1-\tau)\delta(x) + \tau \sum_{\ell = 1}^L \omega_\ell\, \mc{N}_+(x;\theta_\ell,\phi_\ell),
\label{eq:mmsepri}
\end{equation}
where $\mc{N}_+(\cdot)$ denotes the non-negative Gaussian pdf, 
\begin{equation}
\mc{N}_+(x;\theta,\phi) = 
\begin{cases}
\frac{\mc{N}(x;\theta,\phi)}{\Phi_c(-\theta/\sqrt{\phi})} &x \geq 0 \\
0 & x < 0
\end{cases},
\label{eq:nng}
\end{equation}
$\tau\in(0,1]$ is the sparsity rate, and $\omega_\ell,\theta_\ell$, and $\phi_\ell$ are the weight, location, and scale, respectively, of the $\ell^{th}$ mixture component.
For now, we treat the NNGM parameters $[\tau, \vec{\omega}, \vec{\theta}, \vec{\phi}]$ and the model order $L$ as fixed and known. 

As shown in \appref{MMSE_NNGM}, the sum-product GAMP quantities in (R13) and (R14) of \tabref{gamp} then become
\begin{align}
\label{eq:mean_final}
\hat{x}_n &= \frac{\tau}{\zeta_n} \sum_{\ell=1}^L \beta_{n,\ell} \big(\gamma_{n,\ell} + \sqrt{\nu_{n,\ell}}h(\alpha_{n,\ell})\big) \\
\label{eq:var_final}
\mu^x_n  &= 
\frac{\tau}{\zeta_n} \!\sum_{\ell =1}^L \beta_{n,\ell} \Big(\!\nu_{n,\ell} g(\alpha_{n,\ell}) \!+\!\! \big(\gamma_{n,\ell} \!+\!\! \sqrt{\nu_{n,\ell}}h(\alpha_{n,\ell})\big)^2\Big) \!\!-\! \hat{x}_n^2,  
\end{align}
where $\zeta_n$ is the normalization factor 
\begin{equation}
\label{eq:norm}
\zeta_n \defn (1-\tau) \mc{N}(0;\rhat,\mur) + \tau \sum_{\ell =1}^L \beta_{n,\ell},
\end{equation}
$h(\cdot)$ and $g(\cdot)$ were defined in \eqref{h} and \eqref{g}, respectively, 
and 
\begin{align}
\label{eq:alpha}
\alpha_{n,\ell} &\defn \frac{-\gamma_{n,\ell}}{\sqrt{\nu_{n,\ell}}}\\
\label{eq:gamma}
\gamma_{n,\ell} &\defn \frac{\rhat /\mur + \theta_\ell/ \phi_\ell}{1/\mur + 1/\phi_\ell},\\ 
\label{eq:nu}
\nu_{n,\ell} &\defn \frac{1}{1/\mur + 1/\phi_\ell}\\
\label{eq:beta}
\beta_{n,\ell} &\defn \frac{\omega_\ell \mc{N}(\rhat;\theta_\ell,\mur \!+\! \phi_\ell)\Phi_c(\alpha_{n,\ell})}{\Phi_c(-\theta_\ell/ \sqrt{\phi_\ell})}.
\end{align}

From \eqref{gamppost} and \eqref{mmsepri}, it follows that GAMP's approximation to the posterior activity probability $\text{Pr}\{\X_n \neq 0\giv\vec{y}\}$ is 
\begin{equation}
\pi_n = \frac{1}{1+\left(\frac{\tau}{1-\tau}\frac{ \sum_{\ell=1}^L \beta_{n,\ell}}{\mc{N}(0; \rhat ,\mur)}\right)^{-1}}.
\label{eq:postact}
\end{equation}


\section{EM learning of the prior parameters} 			\label{sec:EM}

In the sequel, we will use $\vec{q}$ to refer to the collection of prior parameters.
For example, if NNGM-GAMP was used with the AWGN observation model, then $\vec{q}=[\tau, \vec{\omega}, \vec{\theta}, \vec{\phi}, \psi]$.
Since the value of $\vec{q}$ that best fits the true data is typically unknown, we propose to learn it using an EM procedure \cite{Dempster:JRSS:77}.
The EM algorithm is an iterative technique that is guaranteed to converge to a local maximum 
of the likelihood $f(\vec{y};\vec{q})$.

To understand the EM algorithm, it is convenient to write the log-likelihood as \cite{Vila:TSP:13}
\begin{equation}
\label{eq:EM_decomp}
\ln f(\vec{y};\vec{q}) = \mc{Q}_{\hat{p}}(\vec{y};\vec{q}) + D\big(\hat{p}\,||\,f_{\vX|\vY}(\cdot|\vec{y};\vec{q})\big),
\end{equation}
where $\hat{p}$ is an arbitrary distribution on $\vX$, $D\big(\hat{p}\,||\,\hat{q}\big)$ is the Kullback-Leibler (KL) divergence between $\hat{p}$ and $\hat{q}$, and
\begin{equation}
\label{eq:lb}
\mc{Q}_{\hat{p}}(\vec{y};\vec{q}) \defn E_{\hat{p}}\{\ln f_{\vX,\vY}(\vX,\vec{y};\vec{q})\} + H(\hat{p}),
\end{equation}
where $H(\hat{p})$ is the entropy of $\vX\sim \hat{p}$. 
Importantly, the non-negativity of KL divergence implies that $\mc{Q}_{\hat{p}}(\vec{y};\vec{q})$ is a lower bound on \eqref{EM_decomp}. 
Starting from the initialization $\vec{q}^0$, the EM algorithm iteratively improves its estimate $\vec{q}^i$ at each iteration $i\in\Nat$:
first, it assigns $\hat{p}^i(\cdot) = f_{\vX|\vY}(\cdot|\vec{y};\vec{q}^i)$ to tighten the bound, and then it sets $\vec{q}^{i+1}$ to maximize \eqref{lb} with $\hat{p}=\hat{p}^i$.

Since the exact posterior pdf $f_{\vX|\vY}(\cdot|\vec{y};\vec{q}^i)$ is difficult to calculate, in its place we use GAMP's approximate posterior $\prod_n f_{\X_n|\R_n}(\cdot|\hat{r}_n; \mu^r_n;\vec{q}^i)$ from \eqref{gamppost},
resulting in the EM update
\begin{align}
\vec{q}^{i+1}
&=\argmax_{\vec{q}} \hat{\E}\{\ln f(\vX,\vec{y};\vec{q})\giv \vec{y};\vec{q}^i\},
\label{eq:EMmain}
\end{align}
where $\hat{\E}$ denotes expectation using GAMP's approximate posterior.
Also, because calculating the joint update for $\vec{q}$ in \eqref{EMmain} can be difficult, we perform the maximization \eqref{EMmain} one component at a time, known as ``incremental EM'' \cite{Neal:Jordan:99}.
Note that, even when using an approximate posterior and updating incrementally, the EM algorithm iteratively maximizes a lower-bound to the log-likelihood.

Whereas \cite{Vila:TSP:13} proposed the use of \eqref{EMmain} to tune \emph{sum-product} GAMP, where the marginal posteriors $f_{\X_n|\R_n}(\cdot|\hat{r}_n; \mu^r_n;\vec{q}^i)$ from \eqref{gamppost} are computed for use in steps (R13)-(R14) of \tabref{gamp}, we hereby \textb{propose} the use of \eqref{EMmain} to tune \emph{max-sum} GAMP. 
The reasoning behind our proposal goes as follows.
Although max-sum GAMP does not compute marginal posteriors (but rather joint MAP estimates), its large-system-limit analysis (under i.i.d sub-Gaussian $\vec{A}$) \cite{Javanmard:IMA:13} shows that $\hat{r}_n(t)$ can be modeled as an AWGN-corrupted measurement of the true $x_n$ with AWGN variance $\mu_n^r(t)$, revealing the \emph{opportunity} to compute marginal posteriors via \eqref{gamppost} as an additional step.
Doing so enables the use of \eqref{EMmain} to tune max-sum GAMP.


\subsection{EM update of AWGN variance} 																\label{sec:EM_AWGN}

\textb{We first derive the EM update of the AWGN noise variance $\psi$ (recall \eqref{AWGN}).
This derivation differs from the one in \cite{Vila:TSP:13} in that here we use $\vX$ as the hidden variable (rather than $\vZ$), since experimentally we have observed gains in the low-$\SNR$ regime (e.g., $\SNR<10$ dB).
Because we can write $f(\vec{x},\vec{y};\vec{q}) =D\prod_{m=1}^M f_{\Y|\Z}(y_m|\vec{a}\tran_m\vec{x};\psi)$ with a $\psi$-invariant term $D$, the incremental update of $\psi$ from \eqref{EMmain} becomes
\begin{equation}
\psi^{i+1} 
= \argmax_{\psi > 0}\sum_{m=1}^M \hat{\E}\big\{\ln f_{\Y|\Z}(y_m|\vec{a}_m\tran\vX; \psi) \biggiv \vec{y}; \psi^i\big\}.
	\label{eq:MLAWGN}
\end{equation}
In \appref{AWGN_upd}, we show that \eqref{MLAWGN} reduces to
\begin{equation}
\label{eq:EMAWGNfinal}
\psi^{i+1} = \frac{1}{M}\norm{\vec{y} - \vec{A}\hvec{x}}_2^2 + \frac{1}{M}\sum_{m=1}^M\sum_{n=1}^N a^2_{mn} \mu^x_n.
\end{equation}
}

\subsection{EM update of Laplacian rate parameter} 												\label{EM_Lap}

\textb{
As in the AWGN case above, the incremental update of the Laplacian rate $\psi$ from \eqref{EMmain} becomes
\begin{equation}
\psi^{i+1} 
= \argmax_{\psi > 0}\sum_{m=1}^M \hat{\E}\big\{\ln f_{\Y|\Z}(y_m|\vec{a}_m\tran\vX; \psi) \biggiv \vec{y}; \psi^i\big\},
	\label{eq:MLLaprate}
\end{equation}
but where now $f_{\Y|\Z}$ is given by \eqref{LAP}.
In \appref{lap_upd}, we show that \eqref{MLLaprate} reduces to
\begin{align}
\psi^{i+1} &= M \left( \textstyle \sum_{m=1}^M \hat{\E}\big\{ |\vec{a}\tran_m\vX - y_m| \biggiv \vec{y}; \psi^i \big\}    \right)^{-1}    \label{eq:EM_Lap}
\end{align}
where 
\begin{align}
\hat{\E}\big\{ |\vec{a}\tran_m\vX \!-\! y_m | \biggiv \vec{y}; \psi^i\big\} &\approx \Phi_c\left(\frac{\tilde{z}_m}{\mu^p_m}\right)\left(\tilde{z}_m \!+\! \sqrt{\mu^p_m} h\left(\frac{-\tilde{z}_m}{\mu^p_m}\right)\right) \nonumber \\ 
\label{eq:Lapquant}
&-  \Phi_c\left(\frac{-\tilde{z}_m}{\mu^p_m}\right)\left(\tilde{z}_m \!-\! \sqrt{\mu^p_m} h\left(\frac{\tilde{z}_m}{\mu^p_m}\right)\right)
\end{align}
for $\tilde{z}_m \defn \vec{a}\tran_m\hvec{x} - y_m$, $\mu^p_m$ defined in line (R1) of \tabref{gamp}, and $h(\cdot)$ defined in \eqref{h}.
}

\subsection{EM update of exponential rate parameter} 												\label{EM_exp}
\textb{
Noting that $f(\vec{x},\vec{y};\vec{q}) =D\prod_{n=1}^N f_\X(x_n;\chi)$ with $\chi$-invariant $D$, the incremental EM update of the exponential rate parameter $\chi$ is
\begin{align}
{\chi}^{i+1} 
&= \argmax_{\chi >0}\sum_{n=1}^N \hat{\E}\big\{\ln f_\X(\X_n; \chi) \biggiv \vec{y}; \chi^i\big\},
	\label{eq:MLexprate} \\
&= \argmax_{\chi > 0} N \log \chi - \chi \sum_{n=1}^N \hat{\E}\big\{\X_n \biggiv \vec{y}; \chi^i\big\} \label{eq:MLexprate2}
\end{align}
which, after zeroing the derivative of \eqref{MLexprate2} w.r.t.\ $\chi$, reduces to 
\begin{align}
\chi^{i+1} &= N \left( \sum_{n=1}^N \tilde{r}_n + \sqrt{\mu^r_n} \textb{h}\left(-\frac{\tilde{r}_n}{\sqrt{\mu^r_n}}\right)  \right)^{-1}    \label{eq:EM_exp}
\end{align}
for $\tilde{r}_n \defn \hat{r}_n - \chi \mu^r_n$, $\mu^r_n$ defined in line (R9) of \tabref{gamp}, and $h(\cdot)$ defined in \eqref{h}.
}\textb{The derivation of \eqref{EM_exp} uses the fact that the posterior used for the expectation in \eqref{MLexprate2} simplifies to $f_{\X|\R}(x_n| \hat{r}_n; \mu^r_n) = \mc{N}_+(x_n;\tilde{r}_n,\mu^r_n)$.}
Note that this procedure, \textb{when used in conjunction with the AWGN variance learning procedure,} automatically ``tunes'' the LASSO regularization parameter $\lambda$ in \eqref{main}, a 
difficult problem (see, e.g., \cite{Giryes:ACHA:11}).

\subsection{EM updates for NNGM parameters and model-order selection} \label{sec:EM_NNGM}

Noting that $f(\vec{x},\vec{y};\vec{q}) =D\prod_{n=1}^N f_\X(x_n;\vec{\omega},\vec{\theta},\vec{\phi})$ with $[\vec{\omega},\vec{\theta},\vec{\phi}]$-invariant $D$, the incremental EM updates become
\begin{eqnarray}
{\theta}_k^{i+1} 
&=& \argmax_{\theta_k \in \mathbb{R}}\sum_{n=1}^N \hat{\E}\big\{\ln f_\X(\X_n; \theta_k, \vec{q}^i_{\setminus \theta_k}) \biggiv \vec{y}; \vec{q}^i\big\}, 
	\label{eq:MLtheta}\\
{\phi}_k^{i+1} 
&=& \argmax_{\phi_k > 0}\sum_{n=1}^N \hat{\E}\big\{\ln f_\X(\X_n; \phi_k, \vec{q}_{\setminus \phi_k}^i) \biggiv \vec{y}; \vec{q}^i\big\},
	\label{eq:MLphi} \\
{\vec{\omega}}^{i+1} 
&=& \hspace{-3mm}\argmax_{\vec{\omega}>0:\,\sum_k\!\omega_k=1}\sum_{n=1}^N \hat{\E}\big\{\ln f_\X(\X_n; \vec{\omega}, \vec{q}^i_{\setminus \vec{\omega}}) \biggiv \vec{y}; \vec{q}^i\big\}, \quad
	\label{eq:MLomega} 
\end{eqnarray}
where we use ``$\vec{q}^i\remove{\vec{\omega}}$'' to denote the vector $\vec{q}^i$ with $\vec{\omega}$ components removed (and similar for $\vec{q}^i_{\setminus \theta_k}$ and $\vec{q}_{\setminus \phi_k}^i$).
As derived in \appref{NNGM_upd}, the updates above can be approximated as
\begin{align}
 \label{eq:EMthetafinal}
\theta_k^{i+1} &= \frac{\sum_{n=1}^N \pi_n\overline{\beta}_{n,k}\big(\gamma_{n,k} + \sqrt{\nu_{n,k}}h(\alpha_{n,k})\big)}
{\sum_{n=1}^N \pi_n\overline{\beta}_{n,k}} \\
\label{eq:EMSphifinal}
\phi_k^{i+1} &= \frac{\sum_{n=1}^N \pi_n \overline{\beta}_{n,k}\big(\gamma_{n,k} + \sqrt{\nu_{n,k}}h(\alpha_{n,k}) - \theta_k\big)^2}
{\sum_{n=1}^N \pi_n \overline{\beta}_{n,k}}\nonumber \\
&\quad + \frac{\sum_{n=1}^N \pi_n \overline{\beta}_{n,k} \nu_{n,k} g(\alpha_{n,k})}{\sum_{n=1}^N \pi_n \overline{\beta}_{n,k}}\\
\label{eq:EMSomegafinal}
\omega_k^{i+1} &= \frac{ \sum_{n=1}^N \pi_n \overline{\beta}_{n,k} }
{\sum_{n=1}^N  \pi_n },
\end{align}
where the quantities $\alpha_{n,\ell},\gamma_{n,\ell},\nu_{n,\ell},\beta_{n,\ell},\pi_n$ were defined in \eqref{alpha}-\eqref{postact} and $\overline{\beta}_{n,k} \defn \beta_{n,k}/\sum_\ell \beta_{n,\ell}$.  
\textb{The EM update of the NNGM sparsity rate $\tau$ (recall \eqref{mmsepri}) is identical to that for the GM sparsity rate derived in \cite{Vila:TSP:13}:
\begin{align}
\label{eq:EMStaufinal}
\tau^{i+1}
&= \frac{1}{N}\sum_{n=1}^N \pi_n.
\end{align}}
Since \textb{the quantities in \eqref{EMthetafinal}-\eqref{EMStaufinal}} are already computed by NNGM-GAMP, the EM updates do not significantly increase the complexity beyond that of NNGM-GAMP itself.


The number of components $L$ in the NNGM model \eqref{mmsepri} can be selected using the standard penalized log-likelihood approach to model-order-selection \cite{Stoica:SPM:04}, i.e., by maximizing
\begin{align}
\ln f(\vec{y};\hvec{q}_L) - \eta(L) ,	\label{eq:MOS}
\end{align}
where $\hvec{q}_L$ is the ML estimate of $\vec{q}$ under the hypothesis $L$ (for which we would use the EM estimate) and $\eta(L)$ is a penalty term such as that given by the Bayesian information criterion (BIC).
Since this model-order-selection procedure is identical to that proposed for \textb{EM-GM-GAMP} in \cite{Vila:TSP:13}, we refer interested readers to \cite{Vila:TSP:13} for more details.
In practice, we find that the fixed choice of $L =3$ performs sufficiently well (see \secref{results}).

\subsection{EM initialization} \label{sec:EMinits}

With EM, a good initialization is essential to avoiding bad local minima.
For \textb{EM-NNL-GAMP}, we suggest setting the initial exponential rate parameter \textb{$\chi^0 = 10^{-2}$}, as this seems to perform well over a wide range of problems (see \secref{results}).  

For \textb{EM-NNGM-GAMP}, we suggest the initial sparsity rate 
\begin{equation}
\tau^0 = \min\big \{\tfrac{M}{N}\rho_\text{SE}(\tfrac{M}{N}), 1-\epsilon \big\}
					\label{eq:lambda0}
\end{equation}
where $\epsilon >0$ is set arbitrarily small and $\rho_\text{SE}(\cdot)$ is the theoretical noiseless phase-transition-curve (PTC) for $\ell_1$ recovery of sparse non-negative signals, shown in \cite{Donoho:PNAS:09} to have the closed-form expression 
\begin{eqnarray}
\rho_{\textsf{SE}}(\delta)
= \max_{c\geq 0} \displaystyle
	\frac{1-(1/\delta)[(1+c^2)\Phi(-c)-c\,\varphi(c)]}
	{1+c^2-[(1+c^2)\Phi(-c)-c\,\varphi(c)]} 
					\label{eq:PTC}
\end{eqnarray}
where $\Phi(\cdot)$ and $\varphi(\cdot)$ denote the cdf and pdf of the standard normal distribution.
We then propose to set the initial values of the NNGM weights $\{\omega_{\ell}\}$, locations $\{\theta_{\ell}\}$, and scales $\{\phi_{\ell}\}$ at the values that best fit the uniform pdf on $[0,\sqrt{3 \varphi^0}]$, which can be computed offline similar to the standard EM-based approach described in \cite[p.\ 435]{Bishop:Book:07}.
Under the AWGN model \eqref{AWGN}, we propose to set the initial variance of the noise and signal, respectively, as
\begin{equation}
\psi^0 = \frac{ \norm{\vec{y}}_2^2}{(\SNR+1)M}, \
\varphi^0 = \frac{\norm{\vec{y}}_2^2 - M \psi^0}{||\vec{A}||_F^2\tau^0},
					\label{eq:varphi0}
\end{equation}
where, without knowledge of the true $\SNR\defn{\norm{\vec{Ax}}_2^2}/{\norm{\vec{w}}_2^2}$, we suggest using the value $\SNR\!=\!100$.
Meanwhile, under the i.i.d Laplacian noise model \eqref{LAP}, we suggest to initialize the rate as $\psi^0 = 1$ and $\varphi^0$ again as in \eqref{varphi0}.


\section{Numerical Results} 						\label{sec:results}
The subsections below describe numerical experiments used to ascertain the performance of the proposed methods\footnote{We implemented the proposed algorithms using the GAMPmatlab \cite{GAMPmatlab} package available at \url{http://sourceforge.net/projects/gampmatlab/}.} to existing methods for non-negative signal recovery.

\subsection{Validation of \textb{NNLS-GAMP} and \textb{NNL-GAMP}}     	\label{sec:validation}
We first examine the performance of our proposed algorithms on the linearly constrained NNLS problem \eqref{main} with $\lambda \!=\! 0$. 
In particular, we compare the performance of \textb{NNLS-GAMP} to Matlab's solver \texttt{lsqlin}.
To do this, we drew realizations of $K$-sparse simplex $\vec{x}\in\simp$, where the nonzero elements $\{\underline{x}_k\}_{k=1}^{K}$ were placed uniformly at random and drawn from a symmetric Dirichlet distribution with concentration $a$, i.e.,
\begin{subequations} \label{eq:Dirichlet}
\begin{eqnarray}
&&f(\underline{x}_1,\dots,\underline{x}_{K-1}) 
= \begin{cases} \frac{\Gamma(aK)}{\Gamma(a)^K}\prod_{k=1}^K \underline{x}_k^{a-1},
&\underline{x}_k\in [0,1] \\ 
0 & \text{else} \end{cases}\qquad \\
&&f(\underline{x}_K|\underline{x}_1,\dots,\underline{x}_{K-1})
= \delta(1-\underline{x}_1-\dots-\underline{x}_{K}) ,
\end{eqnarray}  
\end{subequations}
where $\Gamma(\cdot)$ is the gamma function.   
For this first experiment, we used $a\!=\!1$, in which case $\{\underline{x}_k\}_{k=1}^{K-1}$ are i.i.d uniform on $[0,1]$, as well as $K=N$ (i.e., no sparsity).
We then constructed noisy measurements $\vec{y}\in\Real^M$ according to \eqref{y} using $\vec{A}$ with i.i.d $\mc{N}(0,M^{-1})$ entries,
$\SNR \defn \norm{\vec{Ax}}_2^2/\norm{\vec{w}}_2^2 = [10,100,1000]$,  
and sampling ratio $M/N = 3$. 
\tabref{dir} reports the resulting \emph{comparative} $\overline{\NMSE} \defn \norm{\hvec{x}_{\textsf{\textb{NNLS-GAMP}}} - \hvec{x}_{\texttt{lsqlin}}}_2^2/\norm{\vec{x}}_2^2$ and runtime averaged over $R=100$ realizations for signal lengths $N = [100,250,500]$.
From the table, we see that \textb{NNLS-GAMP} and \texttt{lsqlin} return identical solutions (up to algorithmic tolerance\footnote{The algorithms under test include user-adjustable stopping tolerances.  As these tolerances are decreased, we observe that the comparative $\overline{\NMSE}$ also decreases, at least down to Matlab's numerical precision limit.}), but that \textb{NNLS-GAMP}'s runtime scales like $O(N^2)$ while \texttt{lsqlin}'s scales like $O(N^3)$, making \textb{NNLS-GAMP} much faster for larger problem dimensions $N$.
Moreover, we see that \textb{NNLS-GAMP}'s runtime is invariant to SNR, whereas \texttt{lsqlin}'s runtime quickly degrades as the SNR decreases.

\putTable{dir}{\textb{NNLS-GAMP} vs. \texttt{lsqlin}: average comparative $\overline{\NMSE}$ [dB] and runtime [sec] for simplex signal recovery.}{
\newcommand\Theight{\rule{0pt}{2.6 ex}} 
\begin{tabular}{|@{\,}c@{\,}|@{\,}l@{\,}|@{}c@{\,}|@{\,}c@{\,}|@{\,}c@{\,}||@{}c@{\,}|@{\,}c@{\,}|@{\,}c@{\,}||@{}c@{\,}|@{\,}c@{\,}|@{\,}c@{\,}|}
\cline{3-11}
\multicolumn{2}{c|}{} &\multicolumn{3}{c||}{$N = 100$} &\multicolumn{3}{|c||}{$N = 250$} &\multicolumn{3}{|c|}{$N = 500$}\\ 
\cline{3-11}
\multicolumn{2}{c|}{} & & \multicolumn{2}{c||}{time} & &\multicolumn{2}{c||}{time} & &\multicolumn{2}{c|}{time}\\ \cline{4-5} \cline{7-8} \cline{10-11}
\multicolumn{2}{c|}{} &\tiny $\overline{\NMSE}$ &\tiny \textsf{\textb{NNLS-GAMP}} &\tiny \texttt{lsqlin} &\tiny $\overline{\NMSE}$
&\tiny \textsf{\textb{NNLS-GAMP}} &\tiny \texttt{lsqlin} &\tiny $\overline{\NMSE}$ &\tiny \textsf{\textb{NNLS-GAMP}} &\tiny \texttt{lsqlin}\\ \cline{3-11}
\cline{3-11} \hline
\multirow{3}{*}{\begin{sideways}{$\SNR$}\end{sideways}}
&$10$ &-161.8 &0.068 &\textbf{0.050} &-161.8 &\textbf{0.080} &0.550 &-161.8 &\textbf{0.159} &5.414\\  \cline{2-11}
&$100$ &-161.7 &0.069 &\textbf{0.021} &-154.3 &\textbf{0.080} &0.205 &-161.5 &\textbf{0.154} &1.497\\  \cline{2-11}
&$1000$ &-162.1 &0.068 &\textbf{0.011} &-161.7 &0.079 &\textbf{0.074} &-161.5 &\textbf{0.151} &0.504\\ \hline
\end{tabular}
}

Next, we examine the performance of our proposed algorithms on the non-negative
LASSO problem \eqref{main} with $\lambda>0$. 
In particular, we compare \textb{NNL-GAMP} to TFOCS\footnote{We used Matlab code from \url{http://cvxr.com/tfocs/download/}.} \cite{Becker:MPC:11}. 
For this, $K$-sparse non-negative $\vec{x}$ and noisy observations $\vec{y}$ were constructed as before, but now with $M \!=\! 1000$, $N \!=\! 500$, $K\!<\!N$, and $\SNR \!=\! 20$~dB.
\tabref{NNLAMP} shows the runtimes and comparative $\overline{\NMSE}$ between \textb{NNL-GAMP} and TFOCS for various combinations of sparsity $K$ and regularization weight $\lambda$.
\tabref{NNLAMP} shows that the solutions returned by the two algorithms were identical (up to algorithmic tolerance) but that \textb{NNL-GAMP} ran about $4$ to $8$ times faster than TFOCS.

\putTable{NNLAMP}{\textb{NNL-GAMP} vs. TFOCS: average comparative $\overline{\NMSE}$ [dB] and runtime [sec] for $K$-sparse non-negative signal recovery.}{
\newcommand\Theight{\rule{0pt}{2.6 ex}} 
\begin{tabular}{|@{\,}c@{\,}|@{\,}l@{\,}|@{}c@{\,}|@{\,}c@{\,}|@{\,}c@{\,}||@{}c@{\,}|@{\,}c@{\,}|@{\,}c@{\,}||@{}c@{\,}|@{\,}c@{\,}|@{\,}c@{\,}|}
\cline{3-11}
\multicolumn{2}{c|}{} &\multicolumn{3}{c||}{$K = 50$} &\multicolumn{3}{|c||}{$K = 100$} &\multicolumn{3}{|c|}{$K = 150$}\\ 
\cline{3-11}
\multicolumn{2}{c|}{} & & \multicolumn{2}{c||}{time} & &\multicolumn{2}{c||}{time} & &\multicolumn{2}{c|}{time}\\ \cline{4-5} \cline{7-8} \cline{10-11}
\multicolumn{2}{c|}{} &\tiny $\overline{\NMSE}$ &\tiny \textsf{\textb{NNL-GAMP}} &\tiny TFOCS &\tiny $\overline{\NMSE}$
&\tiny \textsf{\textb{NNL-GAMP}} &\tiny TFOCS &\tiny $\overline{\NMSE}$ &\tiny \textsf{\textb{NNL-GAMP}} &\tiny TFOCS\\ \cline{3-11}
\cline{3-11} \hline
\multirow{3}{*}{\begin{sideways}{$\lambda$}\end{sideways}}
&$0.01$ &-135.7 &\textbf{0.024} &0.091 &-139.9 &\textbf{0.025} &0.119 &-140.8 &\textbf{0.025} &0.104\\  \cline{2-11}
&$0.001$ &-125.4 &\textbf{0.026} &0.130 &-122.9 &\textbf{0.026} &0.148 &-117.0 &\textbf{0.027} &0.175\\  \cline{2-11}
&$0.0001$ &-113.2 &\textbf{0.035} &0.256 &-113.4 &\textbf{0.036} &0.262 &-112.4 &\textbf{0.036} &0.292\\ \hline
\end{tabular}
}

\subsection{Noiseless Empirical Phase Transitions} \label{sec:PTCS}

It has been established (see, e.g., \cite{Donoho:PNAS:09}) that, for the recovery of a non-negative $K$-sparse signal $\vec{x}\in\Real^N$ from noiseless observations $\vec{y}\!=\!\vec{Ax}\in\Real^M$, there exists a sharp phase-transition separating problem sizes $(M,N,K)$ that are perfectly solvable (with very high probability) from those that are not.
The precise location of the phase-transition curve (PTC) differs among algorithms, presenting an avenue for comparison. 

Below, we present empirical PTCs for the recovery of $K$-sparse $N$-length simplex signals from $M$ noiseless measurements. 
To compute each PTC, we fixed $N\!=\!500$ and constructed a $20\times 20$ uniformly spaced grid on the $\frac{M}{N}$-versus-$\frac{K}{M}$ plane for $\frac{M}{N} \in[0.05,1]$ and $\frac{K}{M}\in[0.05,1]$.
At each grid point, we drew $R\!=\!100$ independent realizations of the pair $(\vec{A},\vec{x})$, where $\vec{A}$ was drawn from i.i.d $\mc{N}(0,M^{-1})$ entries and $\vec{x}\in\Real^N$ had $K$ nonzero elements $\{\underline{x}_k\}_{k=1}^{K}$ (placed uniformly at random) drawn from a symmetric Dirichlet distribution \eqref{Dirichlet} with concentration parameter $a$.
For the $r^{th}$ realization of $(\vec{A},\vec{x})$, 
we attempted to recover non-negative sparse $\vec{x}$ from the augmented observations 
$\smallmat{\vec{y}\\1}\!=\!\smallmat{\vec{A}\\\vec{1}\tran}\vec{x}$, which implicitly enforce the simplex constraint. 
The resulting recovery $\hvec{x}$ was considered to be ``successful'' if $\NMSE \!\defn\! \norm{\vec{x}-\hvec{x}}_2^2/\norm{\vec{x}}_2^2 < 10^{-6}$. 
Using $S_r \!=\! 1$ to record a success and $S_r\!=\!0$ a failure, the average success rate was then computed as $\overline{S} \!\defn\! \frac{1}{R}\sum_{r=1}^R S_r$, and the corresponding empirical PTC was plotted as the $\overline{S}\!=\!0.5$ level-curve using Matlab's \texttt{contour} command. 

Figures~\ref{fig:PTC_alpha1}~and~\ref{fig:PTC_alpha100} show the empirical PTCs under the Dirichlet concentration $a \!=\! 1$ (\ie i.i.d uniform $\{\underline{x}_k\}_{k=1}^{K-1}$) and $a \!=\! 100$ (\ie $\underline{x}_k\approx \frac{1}{K}~\forall k$), respectively, for our proposed EM-tuned \textb{NNGM-GAMP} and \textb{NNL-GAMP} algorithms, in comparison to the GSSP\footnote{For GSSP, we used code provided by its authors, but found that its performance was greatly enhanced by initializing the algorithm at the Basis Pursuit solution (as computed by SPGL1 \cite{vandenBerg:JSC:08}) and using the stepsize $100/\norm{\vec{A}}^2_F$.} approach \eqref{GSSP} proposed in \cite{Kyrillidis:ICML:13}.
We did not consider \textb{NNLS-GAMP} and \texttt{lsqlin} because, for $\vec{A}$ drawn i.i.d Gaussian, the solution to the non-negative LS problem ``$\argmin_{\vec{x}\geq\vec{0}}\|\vec{y}-\vec{Ax}\|_2^2$'' is not guaranteed to be unique when $M\!<\!N$ \cite[Thm.~1]{Slawski:EJS:13}, which is the setting considered here. 
Figures~\ref{fig:PTC_alpha1}~and~\ref{fig:PTC_alpha100} also show $\rho_{\textsf{SE}}(\frac{M}{N})$ from \eqref{PTC}, i.e., the theoretical large-system-limit PTC for $\ell_1$-based recovery of sparse non-negative (SNN) signals.

\putFrag{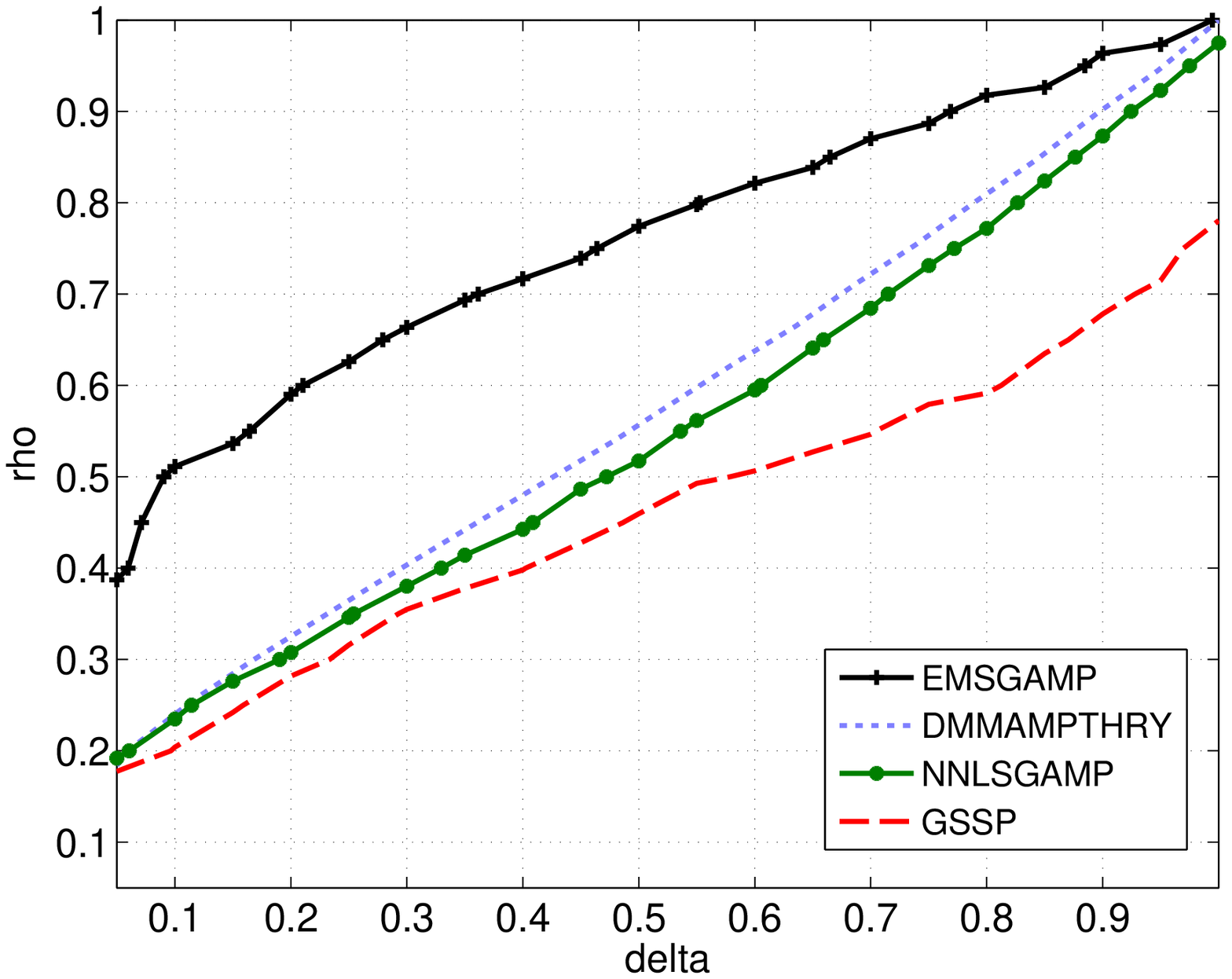}
	{Empirical PTCs and $\ell_1$-SNN theoretical PTC for noiseless recovery of length-$N\!=\!500$, $K$-sparse, simplex signals with Dirichlet concentration $a = 1$ from $M$ measurements.}
	{\figsize}
	{\psfrag{delta}[t][t][0.9]{$M/N$} 
	 \psfrag{rho}[][][0.9]{$K/M$} 
         \newcommand{\sz}{0.51}
         \psfrag{EMSGAMP}[l][l][\sz]{\sf \textb{EM-NNGM-GAMP}}
	 \psfrag{USGAMP}[l][l][\sz]{\sf \textb{NNLS-GAMP}}
	 \psfrag{NNLSGAMP}[l][l][\sz]{\sf \textb{EM-NNL-GAMP}}
         \psfrag{GSSP}[l][l][\sz]{\sf GSSP}
         \psfrag{FCLS}[l][l][\sz]{\tt lsqnonneg}
         \psfrag{DMMAMPTHRY}[l][l][\sz]{\sf theoretical $\ell_1$}}

\putFrag{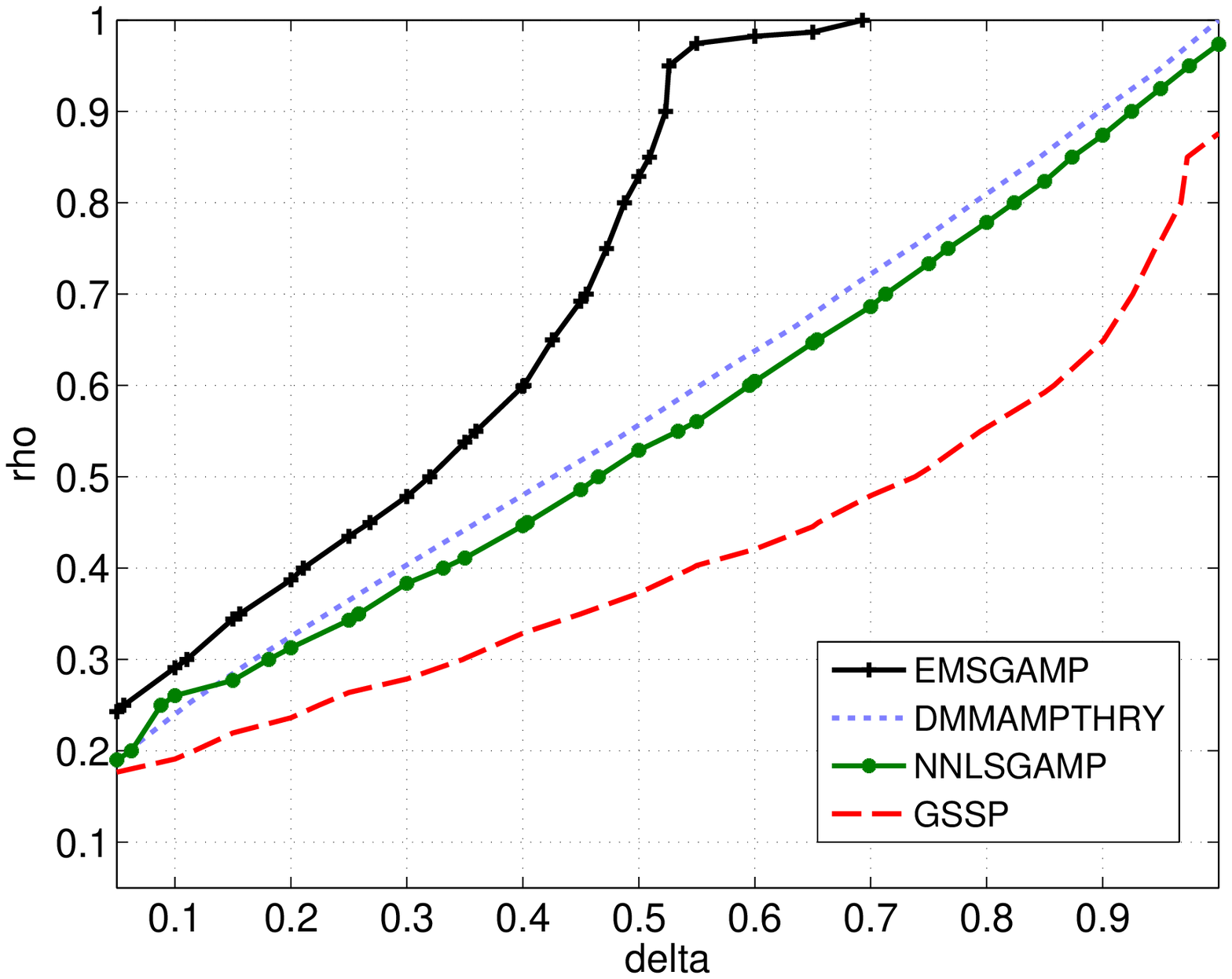}
	{Empirical PTCs and $\ell_1$-SNN theoretical PTC for noiseless recovery of length-$N\!=\!500$, $K$-sparse, simplex signals with Dirichlet concentration $a = 100$ from $M$ measurements.}
	{\figsize}
	{\psfrag{delta}[t][t][0.9]{$M/N$} 
	 \psfrag{rho}[][][0.9]{$K/M$} 
         \newcommand{\sz}{0.51}
         \psfrag{EMSGAMP}[l][l][\sz]{\sf \textb{EM-NNGM-GAMP}}
	 \psfrag{USGAMP}[l][l][\sz]{\sf \textb{NNLS-GAMP}}
	 \psfrag{NNLSGAMP}[l][l][\sz]{\sf \textb{EM-NNL-GAMP}}
         \psfrag{GSSP}[l][l][\sz]{\sf GSSP}
         \psfrag{FCLS}[l][l][\sz]{\tt lsqnonneg}
         \psfrag{DMMAMPTHRY}[l][l][\sz]{\sf theoretical $\ell_1$}}

Looking at Figures~\ref{fig:PTC_alpha1}~and~\ref{fig:PTC_alpha100}, we see that the empirical PTCs of \textb{EM-NNL-GAMP} are close to the theoretical $\ell_1$ PTC, as expected, and significantly better than those of GSSP.
More striking is the far superior PTCs of \textb{EM-NNGM-GAMP}.
We attribute \textb{EM-NNGM-GAMP}'s success to three factors: i) the generality of the NNGM prior \eqref{mmsepri}, ii) the ability of the proposed EM approach to accurately learn the prior parameters, and iii) the ability of sum-product GAMP to exploit the learned prior.
In fact, \figref{PTC_alpha100} shows \textb{EM-NNGM-GAMP} reliably reconstructing $K$-sparse signals from only $M\!=\!K$ measurements in the compressive (i.e., $M<N$) regime. 

\subsection{Sparse Non-negative Compressive Imaging} 			\label{sec:image}

\begin{figure}
\vspace{1 mm}
\includegraphics[width=1.7in]{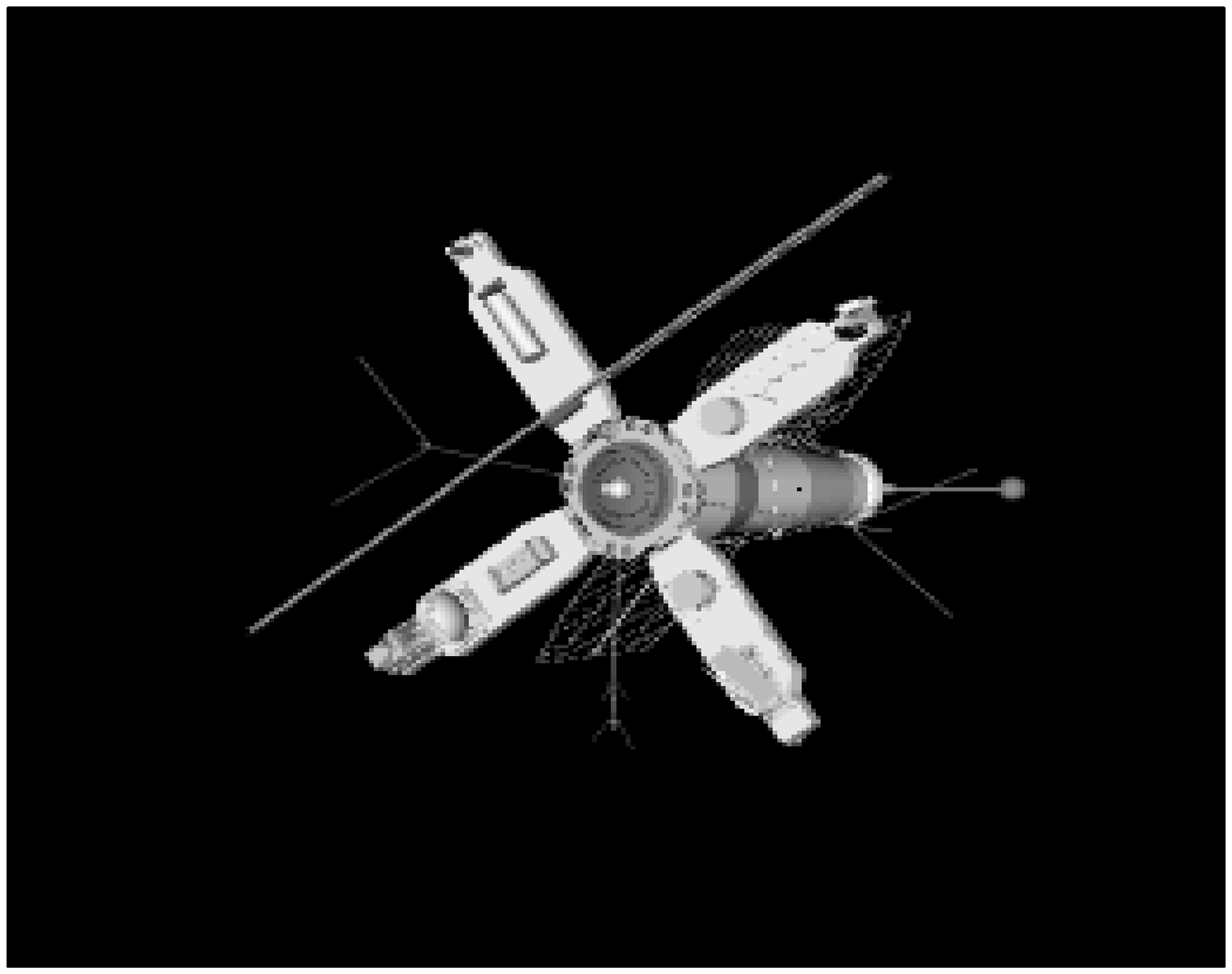}
\hfill
\includegraphics[width=1.7in]{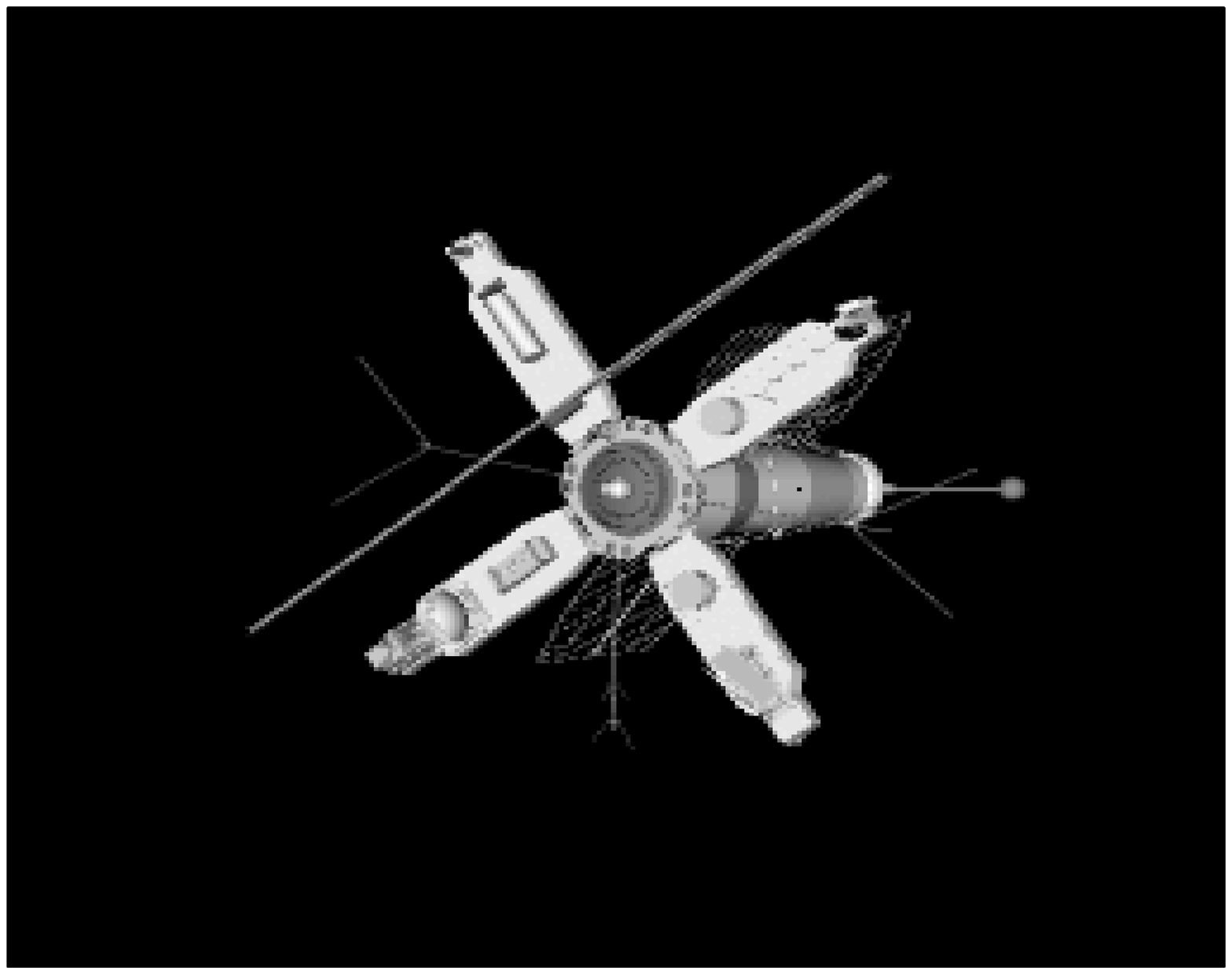}
\caption{Sparse non-negative image of a satellite: original image on left and \textb{EM-NNGM-GAMP} recovery at $\frac{M}{N}=\frac{1}{4}$ on right.}
\label{fig:satellite}
\vspace{-2 mm}
\end{figure}

As a practical example, we experimented with the recovery of a sparse non-negative image. 
For this, we used the $N \!=\! 256\times 256$ satellite image shown on the left of \figref{satellite}, which contained $K \!=\! 6678$ nonzero pixels and $N\!-\!K\!=\!58858$ zero-valued pixels, and thus was approximately $10\%$ sparse. 
Measurements $\vec{y}=\vec{Ax}+\vec{w}\in\Real^M$ were collected under i.i.d Gaussian noise $\vec{w}$ whose variance was selected to achieve an $\SNR \!=\! 60$ dB.
Here, $\vec{x}$ represents the (rasterized) image and $\vec{A}$ a linear measurement operator configured as $\vec{A} \!=\! \vec{\Phi\Psi S}$, where $\vec{\Phi}\in\{0,1\}^{M\times N}$ was constructed from rows of the $N\!\times\! N$ identity matrix selected uniformly at random, $\vec{\Psi}\in\{-1,1\}^{N\times N}$ was a Hadamard transform, and $\vec{S}\in\Real^{N\times N}$ was a diagonal matrix with $\pm 1$ diagonal entries chosen uniformly at random.
Note that multiplication by $\vec{A}$ can be executed using a fast binary algorithm, making it attractive for hardware implementation.
For this experiment, no linear equality constraints exist and so the observation model was not augmented as in \eqref{model}.

As a function of the sampling ratio $\tfrac{M}{N}$, \figref{satimagePTC} shows the $\NMSE$ and runtime averaged over $R \!=\! 100$ realizations of $\vec{A}$ and $\vec{w}$ for the proposed \textb{EM-NNGM-GAMP} and \textb{EM-NNL-GAMP} in comparison to 
\textb{EM-GM-GAMP} from \cite{Vila:TSP:13},
genie-tuned non-negative LASSO via TFOCS \cite{Becker:MPC:11},\footnote{%
Using \textb{EM-NNL-GAMP}'s $\hvec{x}$, we ran TFOCS over an $11$-point grid of hypothesized non-negative $\ell_1$ penalty $\lambda \in \{0.5\norm{\vec{A}\tran(\vec{y} - \vec{A}\hvec{x})}_\infty,\dots,2\norm{\vec{A}\tran(\vec{y} - \vec{A}\hvec{x})}_\infty\}$ and then reported the total runtime and best \NMSE.}
and genie-tuned standard LASSO implemented via SPGL1\footnote{%
We ran SPGL1 in ``BPDN mode,'' \ie solving $\min_{\vec{x}} \norm{\vec{x}}_1$ s.t. $\norm{\vec{y} - \vec{Ax}}_2 < \sigma$ for hypothesized tolerances $\sigma^2 \in \{0.3,0.6,\dots,1.5\} \times M\psi$ and then reported the total runtime and best $\NMSE$.} \cite{vandenBerg:JSC:08}.
NNLS methods were not considered because of the non-uniqueness of their solutions in the $M\!<\!N$ regime (recall \cite[Thm.~1]{Slawski:EJS:13}).

\Figref{satimagePTC} shows that the proposed \textb{EM-NNGM-GAMP} algorithm provided the most accurate signal recoveries for all undersampling ratios.
Remarkably, its phase-transition occurred at $\frac{M}{N}\approx 0.25$, whereas that of the other algorithms occurred at $\frac{M}{N}\approx 0.35$.
The gain of \textb{EM-NNGM-GAMP} over \textb{EM-GM-GAMP} can be attributed to the former's exploitation of signal non-negativity, whereas the gain of \textb{EM-NNGM-GAMP} over non-negative LASSO (either via \textb{EM-NNL-GAMP} or genie-tuned TFOCS) can be attributed to former's learning/exploitation of the true signal distribution.
Finally, the gain of non-negative LASSO over standard LASSO can be attributed to the former's exploitation of signal non-negativity.

\putFrag{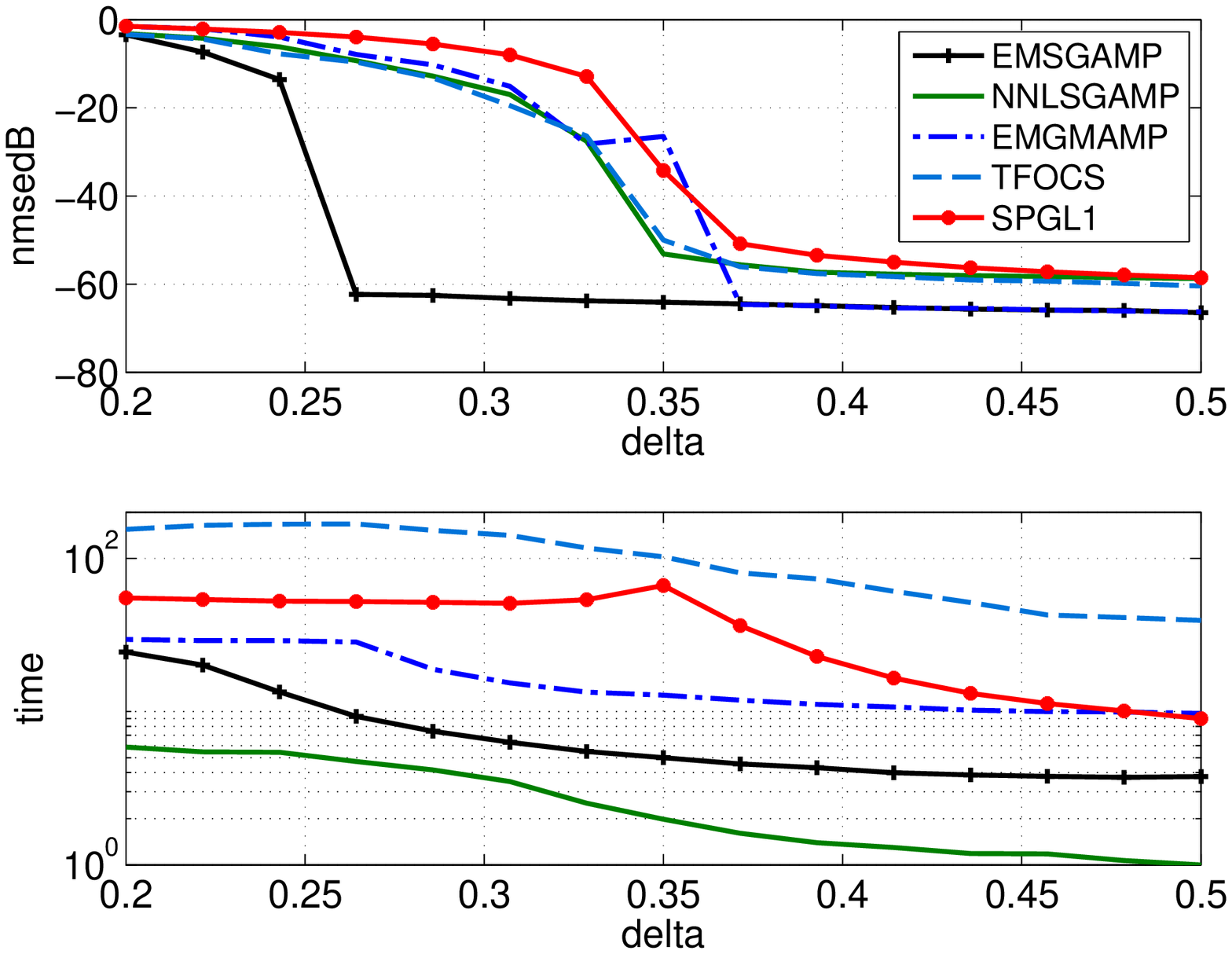}
{Recovery $\NMSE$ (top) and runtime (bottom) versus $\tfrac{M}{N}$ for the sparse NN satellite image for the proposed \textb{EM-NNGM-GAMP} and \textb{EM-NNL-GAMP} compared to \textb{EM-GM-GAMP}, non-negative LASSO via oracle-tuned TFOCS, and standard LASSO via oracle-tuned SPGL1.}
{\figsize}
{\newcommand{\sz}{0.38}
  \psfrag{EMSGAMP}[l][l][\sz]{\sf \textb{EM-NNGM-GAMP}}
  \psfrag{EMGMAMP}[l][l][\sz]{\sf \textb{EM-GM-GAMP}}
   \psfrag{SPGL1}[l][l][\sz]{\sf genie SPGL1}
  \psfrag{NNLSGAMP}[l][l][\sz]{\sf \textb{EM-NNL-GAMP}}
  \psfrag{TFOCS}[l][l][\sz]{\sf genie TFOCS}
  \newcommand{\szz}{0.8}
   \psfrag{delta}[t][c][\szz]{$M/N$}
  \psfrag{time}[c][c][\szz]{\sf time \ [sec]}
   \psfrag{nmsedB}[c][c][\szz]{\sf \NMSE \ [dB]}
}

\Figref{satimagePTC} also demonstrates that the LASSO tuning procedure proposed in \secref{EM} works very well: the $\NMSE$ of \textb{EM-NNL-GAMP} is nearly identical to that of oracle-tuned TFOCS for all sampling ratios $M/N$.

Finally, \figref{satimagePTC} shows that \textb{EM-NNGM-GAMP} was about $3$ times as fast as \textb{EM-GM-GAMP}, between $3$~to~$15$ times as fast as SPGL1 (implementing standard LASSO), and between $10$~to~$20$ times as fast as TFOCS (implementing non-negative LASSO).
The proposed \textb{EM-NNL-GAMP} was about $2$~to~$4$ faster than \textb{EM-NNGM-GAMP}, although it did not perform as well in terms of $\NMSE$.


\subsection{Portfolio Optimization}  									\label{sec:port}
As another practical example, we consider portfolio optimization under the return-adjusted Markowitz mean-variances (MV) framework \cite{Markowitz:Book:91}:
if $\vec{x}\!\in\!\simp$ is a portfolio and $\vR_{M+1}\in\Real^N$ is a random vector that models the returns of $N$ commodities at the future time $M\!+\!1$, then we desire to design $\vec{x}$ so that the future sum-return $\vR_{M+1}\tran\vec{x}$ has relatively high mean and low variance.
Although $\vR_{M+1}$ is unknown at design time, we assume knowledge of the past $M$ returns $\vec{A} \!\defn\! [\vec{r}_1,\dots,\vec{r}_M]\tran$, which can be time-averaged to yield $\vec{\mu} \!\defn\! \tfrac{1}{M} \sum_{m=1}^M \vec{r}_m \!=\!  \tfrac{1}{M}\vec{A}\tran\vec{1}$, and then (assuming stationarity) design $\vec{x}$ that minimizes the variance around a target sum-return of $\rho$, i.e.,
\begin{equation}
\label{eq:Mark}
\hvec{x} = \argmin_{\vec{x} \in \simp} \norm{\vec{1}\rho-\vec{Ax}}_2^2 + \lambda \|\vec{x}\|_1 ~\st~ \vec{\mu}\tran\vec{x} = \rho.
\end{equation}
In \eqref{Mark}, the use of sparsity promoting $\ell_1$ regularization \cite{Brodie:09} aims to help the portfolio designed from past data $\{\vec{r}_m\}_{m=1}^M$ generalize to the future data $\vR_{M+1}$.
Without $\ell_1$ regularization, the solutions to \eqref{Mark} are often outperformed by the ``\naive'' portfolio $\vec{x}_{\textsf{\naive}} \!\defn\! \tfrac{1}{N}\vec{1}$ in practice \cite{Demiguel:09}.
%
%

Noting that \eqref{Mark} is a special case of \eqref{main}, MV portfolio optimization is a natural application for the algorithms developed in this paper.
We thus tested our proposed algorithms against\footnote{We were not able to configure GSSP in a way that maintained $\vec{\mu}\tran \hvec{x}=\rho$, even approximately, after the simplex projection step in \eqref{GSSP}.}
\texttt{lsqlin} and cross-validated (CV)\footnote{For CV-TFOCS, we used $4$-fold cross-validation to tune $\lambda$ over a $15$-point grid between $0.001$ and $0.1$.} TFOCS using the FF$49$ portfolio database,\footnote{The FF49 database and other financial datasets can be obtained from \url{http://mba.tuck.dartmouth.edu/pages/faculty/ken.french/data_library.html}.} which consists of monthly returns for $N \!=\!49$ securities from July 1971 (i.e., $\vec{r}_1$) to July 2011 (i.e., $\vec{r}_{481}$).
In particular, starting from July 1981 and moving forward in yearly increments, we collected the past $M\!=\!120$ months of return data in $\vec{A}(i)\!\defn\! [\vec{r}_{12(i-1)+1},\dots,\vec{r}_{12(i-1) + M}]\tran$ and computed the corresponding time-average return $\vec{\mu}(i)\!\defn\! \tfrac{1}{M} \vec{A}(i)\tran\vec{1}$, where $i\in\{1,\dots,30\}$ indexed the years from 1981 to 2010.
Then, we chose the target sum-return $\rho(i)$ to be that of the \naive\ scheme, i.e., $\rho(i)\!=\!\frac{1}{N}\vec{\mu}(i)\tran\vec{1}$, 
and computed the portfolio $\hvec{x}(i)$ from $\{\vec{A}(i),\vec{\mu}(i),\rho(i)\}$ for each algorithm under test. 
The resulting $\hvec{x}(i)$ was evaluated on the \emph{future} $T\!=\!12$ months of return data using the Sharpe ratio 
$\textsf{SR}(i)\defn \hat{\rho}(i)/\hat{\sigma}(i)$, where
\begin{align}
\hat{\rho}(i) 
&\defn \frac{1}{T}\sum_{t=1}^{T} \vec{r}_{12(i-1)+M+t}\tran \hvec{x}(i),
\label{eq:stats1} \\
\hat{\sigma}^2(i) 
&\defn \frac{1}{T}\sum_{t=1}^{T} \big(\vec{r}_{12(i-1)+M+t}\tran \hvec{x}(i) - \hat{\rho}(i)\big)^2, 
\end{align}

For \texttt{lsqlin}, the constraints were specified directly.
For \textb{NNLS-GAMP}, \textb{EM-NNL-GAMP}, and \textb{EM-NNGM-GAMP}, the constraints were enforced using \eqref{model} with $\vec{B}=[\vec{\mu},\vec{1}]\tran$ and $\vec{c}=[\rho, 1]\tran$, and
for CV-TFOCS, the constraints were enforced using the augmentation
\begin{equation}
\ovec{y} \defn \smallmat{\rho(i)\vec{1}\\ 500\rho(i)\\ 500}~~\text{and}~~\ovec{A} = \smallmat{\vec{A}(i)\\ 500\vec{\mu}(i)\tran\\ 500\,\vec{1}\tran},
\end{equation}
where the gain of $500$ helped to weight the constraints above the loss.
Lastly, we tried our GAMP-based approaches using both the AWGN likelihood \eqref{AWGN} as well as the AWLN likelihood \eqref{LAP}.

\tabref{portfolio} reports the average Sharpe ratios $\textsf{SR}\!\defn\!\frac{1}{30}\sum_{i=1}^{30}\textsf{SR}(i)$ and runtimes for each algorithm under test.
In addition, it reports the average squared constraint error $\mc{E} \!\defn\! \frac{1}{30}\sum_{i=1}^{30}|\vec{\mu}(i)\tran\hvec{x}(i) - \rho(i)|^2$, showing that all algorithms near-perfectly met the target sum-return constraint $\vec{\mu}(i)\tran\hvec{x}(i)\!=\!\rho(i)$.
The table shows that Matlab's \texttt{lsqlin} and AWGN \textb{NNLS-GAMP} (which solve the same NNLS problem) yielded identical Sharpe ratios, which were $\approx 19\%$ larger than the \naive\ value.
Meanwhile, CV-TFOCS and AWGN \textb{EM-NNL-GAMP} (which solve the same NN LASSO problem) yielded very similar Sharpe ratios, also $\approx 19\%$ larger than the \naive\ value. 
As in previous experiments, AWGN \textb{EM-NNGM-GAMP} outperformed both NNLS and NN LASSO, in this case improving on the \naive\ Sharpe ratio by \textb{$24\%$}.
The table also shows that the use of an AWLN likelihood (robust to outliers \cite{Bloomfield:Book:84})
resulted in across-the-board improvements in Sharpe ratio. 
Among the algorithms under test, AWLN \textb{EM-NNGM-GAMP} yielded the best performance, improving the \naive\ Sharpe ratio by \textb{$27\%$}.

In terms of runtimes, Matlab's \texttt{lsqlin} was by far the fastest algorithm, CV-TFOCS was by far the slowest, and the AMP approaches were in-between.  
\textb{NNLS-GAMP} and \textb{NNL-GAMP} were slower here than in \tabref{dir} and \tabref{NNLAMP} because the matrix $\vec{A}$ in this financial experiment had correlated columns and thus required the use of a stronger damping factor in the GAMPmatlab implementation \cite{GAMPmatlab}.

\putTable{portfolio}{Average Sharpe ratio $\textsf{SR}$, constraint error $\mc{E}$ (in dB), and runtime (in sec) versus algorithm for the FF49 dataset.}{
\begin{center}
\begin{tabular}{|c|l|c|c|c|}
\cline{3-5}
\multicolumn{2}{c|}{}  &$\textsf{SR}$ &time (sec) &$\mc{E}$ (dB) \\ \cline{2-5}
\multicolumn{1}{c|}{} &\textsf{\naive}   &0.3135 &-  &-$\infty$ \\ \cline{2-5}
\multicolumn{1}{c|}{} &\texttt{lsqlin} &0.3725 &\textbf{0.06} & -$307.4$ \\ \cline{2-5}
\multicolumn{1}{c|}{} &\textsf{CV-TFOCS} &0.3747 &31.92 &-56.9 \\ \hline
\multirow{3}{*}{\begin{sideways} AWGN \end{sideways}}
&\textsf{\textb{NNLS-GAMP}} &\textb{0.3724} &\textb{0.68} &\textb{-72.0}\\ \cline{2-5}
&\textsf{\textb{EM-NNL-GAMP}} &\textb{0.3725} &\textb{1.48} &\textb{-60.9} \\ \cline{2-5}
&\textsf{\textb{EM-NNGM-GAMP}} &\textb{0.3900} &\textb{6.98} &\textb{-41.5}\\ \hline
\multirow{3}{*}{\begin{sideways}AWLN\end{sideways}}
&\textsf{\textb{NNLS-GAMP}} &\textb{0.3818} &\textb{1.80} &\textb{-56.1} \\ \cline{2-5}
&\textsf{\textb{EM-NNL-GAMP}} &\textb{0.3829}  &\textb{5.14} &\textb{-43.2} \\ \cline{2-5}
&\textsf{\textb{EM-NNGM-GAMP}} &\textb{\textbf{0.3995}} &\textb{2.95} &\textb{-42.3} \\ \hline
\end{tabular}
\end{center}
}

\subsection{Hyperspectral Image Inversion} \label{sec:HSI}

As a final practical example, we consider hyperspectral image inversion \cite{Bioucas:JSTAEO:12}.
A hyperspectral image is like a color image, but instead of $3$ spectral bands (red, green, and blue) it contains $M\gg 3$ spectral bands.
With $T \!=\! T_1 \!\times\! T_2$  spatial pixels, such an image can be represented by a matrix $\vec{Y}\in\Real^{M\times T}$ and, under the macroscopic model, ``unmixed'' into
\begin{equation}
\label{eq:LMM}
\vec{Y} = \vec{AX} + \vec{W} 
\end{equation}
where the $n$th column in $\vec{A}\in\Real^{M\times N}$ is the spectral signature (or ``endmember'') of the $n$th material present in the scene, the $n$th row in $\vec{X}\in\Real_{\geq 0}^{N\times T}$ is the spatial abundance of that material, and $\vec{W}$ is additive noise.
The $t$th column of $\vec{X}$, henceforth denoted as $\vec{x}_t$, describes the distribution of materials within the $t$th pixel, and so for a valid distribution we need $\vec{x}_t\!\in\!\simp$.
We will assume that the endmembers $\vec{A}$ have been extracted from $\vec{Y}$ (e.g., via the well known VCA algorithm \cite{Nasc:TGRS:05}) and therefore focus on image inversion, where the goal is to estimate $\vec{X}$ in \eqref{LMM} given $\vec{Y}$ and $\vec{A}$.
In particular, the goal is to estimate a (possibly sparse) simplex-constrained $\vec{x}_t$ from the observation $\vec{y}_t=\vec{A}\vec{x}_t+\vec{w}_t$ at each pixel $t$.

We evaluated algorithm performance using the SHARE~2012 Avon dataset\footnote{The SHARE~2012 Avon dataset can be obtained from \url{http://www.rit.edu/cos/share2012/}.} \cite{Giannandrea:SPIE:13}, which uses $M\!=\!360$ spectral bands, corresponding to wavelengths between $400$ and $2450$~nm,
over a large rural scene. 
To do this, we first cropped down to the scene shown in \figref{RIT1_rgb}, known to consist primarily of pure grass, dry sand, black felt, and white TyVek \cite{Canham:SPIE:13}.
We then extracted the endmembers $\vec{A}$ from $\vec{Y}$ using VCA.
Finally, we estimated the simplex-constrained columns of $\vec{X}$ from $(\vec{Y},\vec{A})$ using \textb{NNLS-GAMP}, EM-NNL-GAMP, \textb{EM-NNGM-GAMP}, \texttt{lsqlin} (known in the hyperspectral literature as ``fully constrained least squares'' \cite{Heinz:TGRS:01}), and GSSP.
For both \textb{EM-NNL-GAMP} and \textb{EM-NNGM-GAMP}, we opted to learn the prior parameters separately for each \emph{row} of $\vec{X}$, since the marginal distributions can be expected to differ across materials.
For GSSP, we assumed that each pixel was at most $K\!=\!3$-sparse and used a step size of $3/\norm{\vec{A}}_F^2$, as these choices seemed to yield the best results.

\putFrag{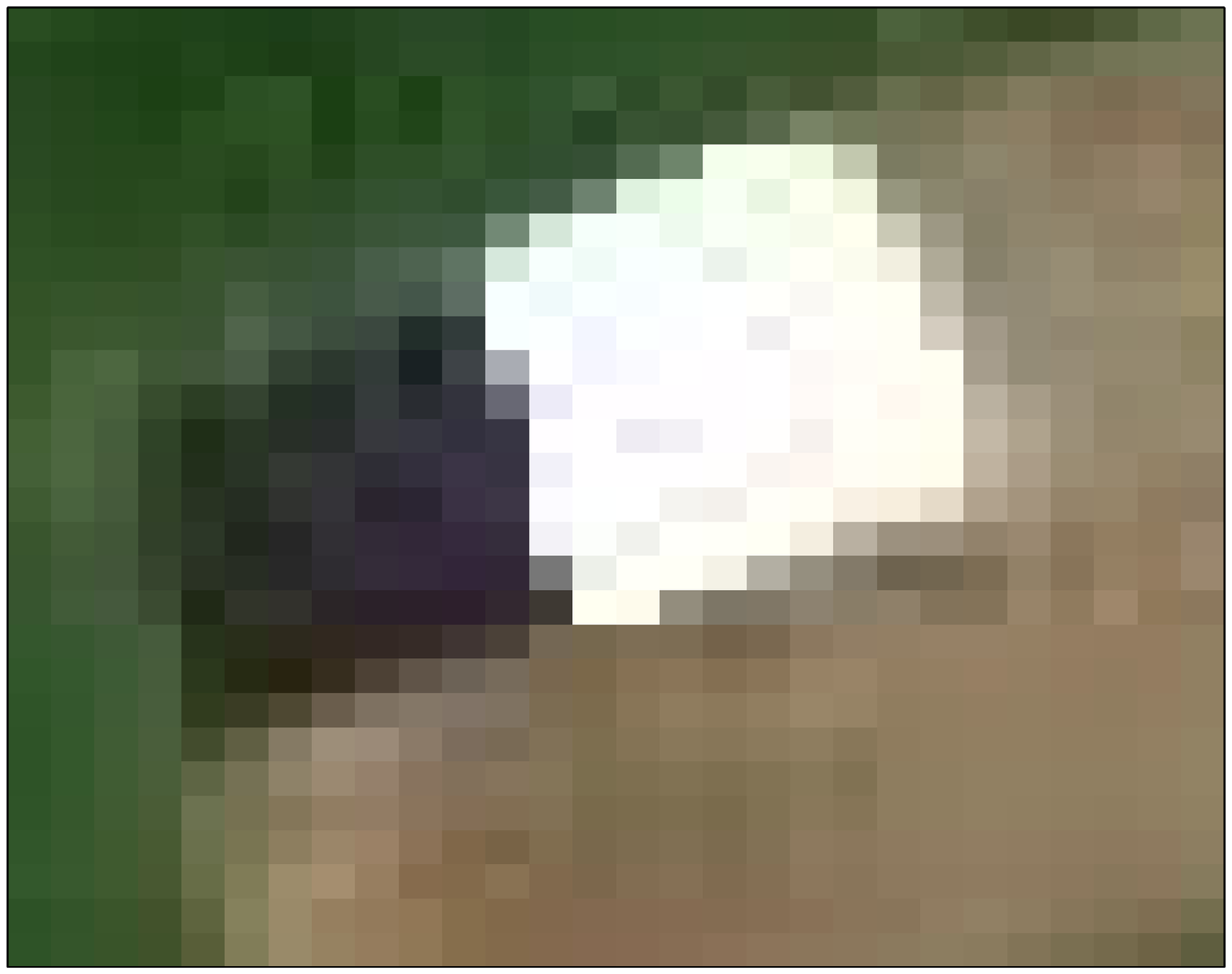}
{RGB image of the cropped scene of the SHARE~2012 dataset \cite{Giannandrea:SPIE:13}.}
{2.3}
{}

\begin{figure}[t]
\newcommand{\pwid}{3.4in}
\newcommand{\fwid}{3.4 in}
  \begin{minipage}{\pwid}
    (a) \texttt{lsqlin} (runtime $= 2.26$ sec):\\
    \includegraphics[width=\fwid]{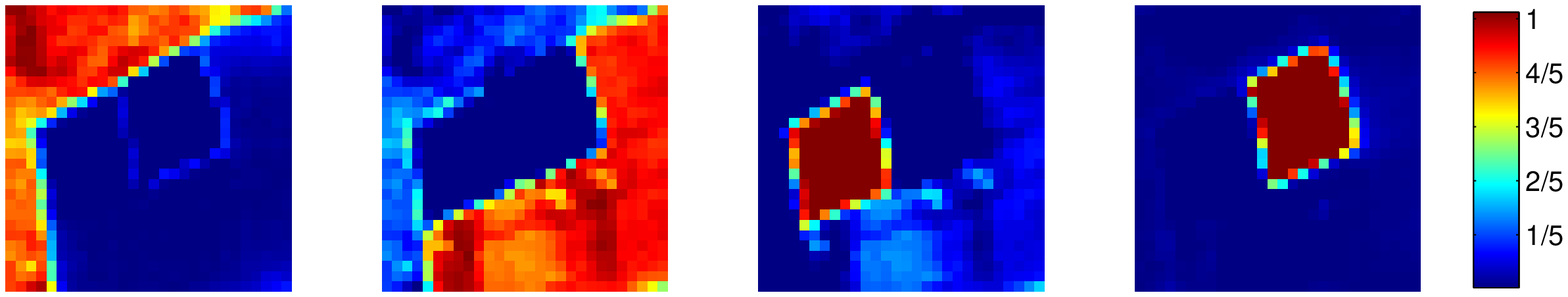}
  \end{minipage} \\[2mm]
  \begin{minipage}{\pwid}
    (b) \textb{NNLS-GAMP} (\textb{runtime $= 2.84$ sec}):\\
    \includegraphics[width=\fwid]{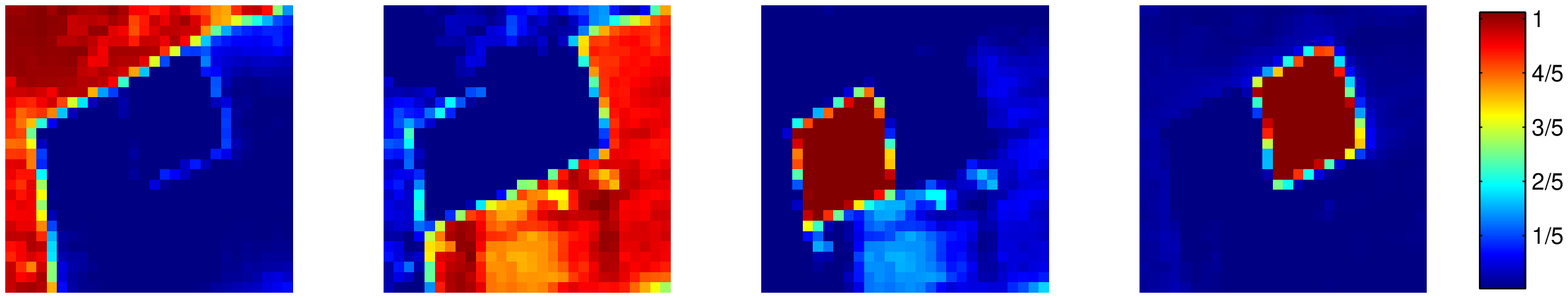}
  \end{minipage} \\[2mm]
  \begin{minipage}{\pwid}
    (c) \textb{EM-NNL-GAMP} (\textb{runtime $= 3.23$ sec}):\\
    \includegraphics[width=\fwid]{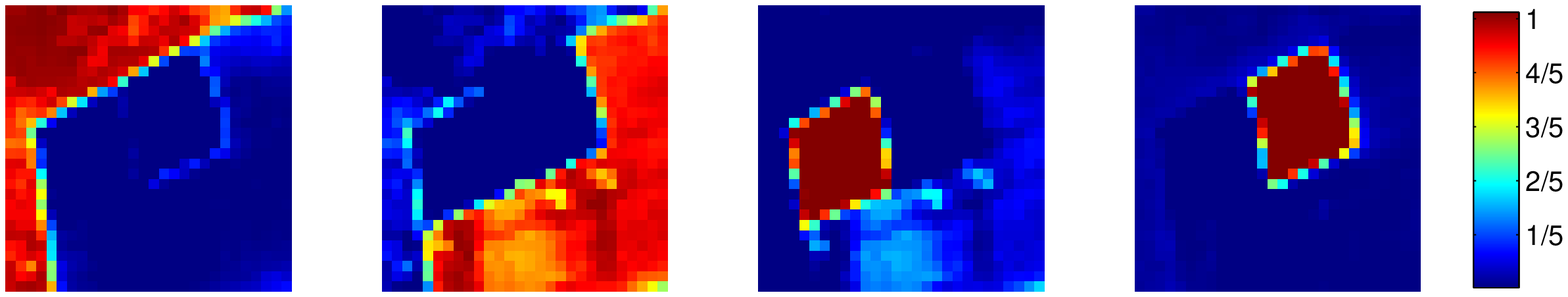}
  \end{minipage} \\[2mm]
  \begin{minipage}{\pwid}
    (d) \textb{EM-NNGM-GAMP} (\textb{runtime $= 4.37$ sec}):\\
    \includegraphics[width=\fwid]{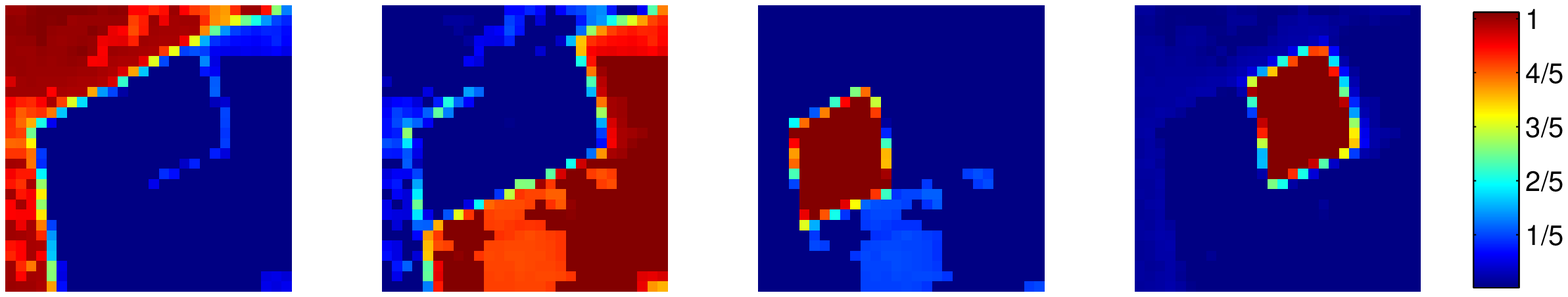}
  \end{minipage} \\[2mm]
  \begin{minipage}{\pwid}
    (e) GSSP (runtime $= 170.71$ sec):\\
    \includegraphics[width=\fwid]{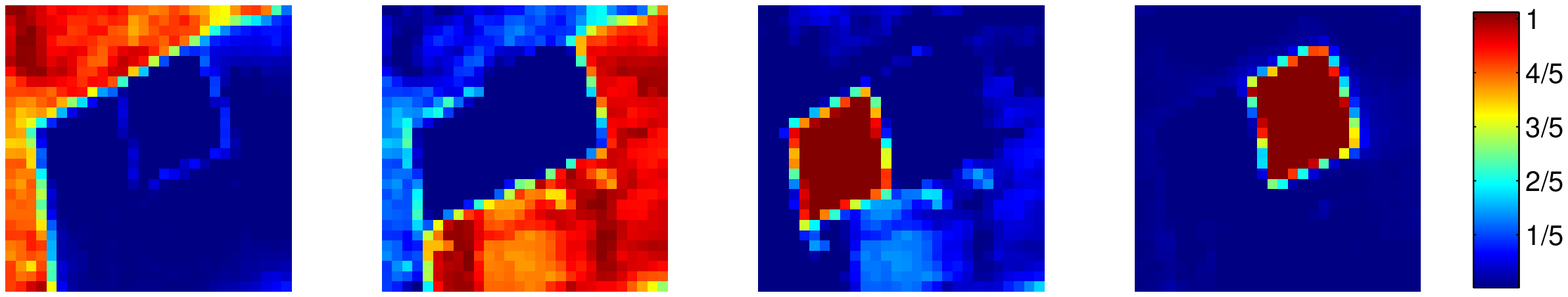}
  \end{minipage}%
  \caption{Each row shows the $N=4$ abundance maps estimated by a given algorithm. From left to right, the materials are: grass, dry sand, black felt, and white TyVek.  \Figref{RIT1_rgb} shows the RGB image of the same scene.}
  \label{fig:abun}
\vspace{-4 mm}
\end{figure}

Since we have no knowledge of the true abundances $\vec{X}$, we are unable to present quantitative results on estimation accuracy.  
However, a qualitative comparison is made possible using the fact that most pixels in this scene are known to be pure \cite{Giannandrea:SPIE:13} (i.e., contain only one material). 
In particular, each row of \figref{abun} shows the $N\!=\!4$ abundance maps recovered by a given algorithm, and we see that all recoveries are nearly pure.
However, the recoveries of \textb{EM-NNGM-GAMP} are the most pure, as evident from the deep blue regions in the first and third columns of \figref{abun}, as well as the deep red regions in the first and second columns.
In terms of runtime, GSSP was by far the slowest algorithm, whereas all the other algorithms were similar (with \texttt{lsqlin} beating the others by a small margin).

\section{Conclusions} 								\label{sec:conc}

The problem of recovering a linearly constrained non-negative sparse signal $\vec{x}$ from noisy linear measurements $\vec{y}$ arises in many applications.
One approach is to pose a sparsity-inducing convex optimization problem like \eqref{main} and then apply standard solvers like \texttt{lsqlin} (when $\lambda=0$) or TFOCS (when $\lambda>0$), although doing so requires also solving the non-trivial problem of optimizing $\lambda$ \cite{Giryes:ACHA:11}.
Another approach is to solve for the MMSE estimate of $\vec{x}$, but doing so is made difficult by the need to estimate the prior distribution of $\vec{x}$ and then compute the resulting posterior mean.

In this paper, we proposed new solvers for \eqref{main} based on the min-sum AMP methodology, yielding \textb{NNLS-GAMP} (for $\lambda=0$) and \textb{NNL-GAMP} (for $\lambda>0$), and we demonstrated computational advantages relative to standard solvers in the large-$N$ regime.
In addition, we proposed a novel EM-based approach to optimizing $\lambda$ that, in our empirical experiments, worked nearly as well as cross-validation and oracle methods.
Moreover, we proposed a new approximate-MMSE estimation scheme that models $\vec{x}$ using an i.i.d Bernoulli non-negative Gaussian-mixture, learns the distributional parameters via the EM algorithm, and exploits the learned distribution via sum-product AMP.
In all of our experiments, the resulting \textb{EM-NNGM-GAMP} algorithm yielded superior performance while maintaining a reasonable computational efficiency.
Finally, for problems where the noise may be non-Gaussian, we developed Laplacian likelihood models for both min-sum and sum-product GAMP, in addition to EM-tuning procedures, and demonstrated performance gains on practical datasets.

\section*{Acknowledgments}
We would like to thank Jason Parker and Justin Ziniel for suggestions on the Laplacian likelihood and \textb{EM-NNL-GAMP}.

\appendices
\section{EM update for AWGN variance} 					\label{app:AWGN_upd}

\textb{
Inserting the Gaussian likelihood \eqref{AWGN} into \eqref{MLAWGN}, we see that the EM update for the noise variance $\psi$ becomes
\begin{equation}
\label{eq:EMAWGN1}
\psi^{i+1} = \argmax_{\psi} \frac{M}{2} \ln \frac{1}{\psi} - \frac{1}{2\psi} \hat{\E}\{ \norm{\vec{y} - \!\vec{A}\vX}_2^2 \biggiv \vec{y}; \psi^i\}, 
\end{equation}
where, for the joint posterior $f_{\vX|\vY}(\vec{x}|\vec{y};\psi^i)$, we use the product of the approximate marginal GAMP posteriors from \eqref{gamppost}.  
By zeroing the derivative of the objective in \eqref{EMAWGN1} w.r.t.\ $\psi$, we find that
\begin{equation}
\label{eq:EMAWGN2}
\psi^{i+1} = \frac{1}{M} \hat{\E}\{ \norm{\vec{y} - \vec{A}\vX}_2^2 \biggiv \vec{y}; \psi^i\},
\end{equation}
where the expectation simplifies to
\begin{align}
\hat{\E}\{ &\norm{\vec{y} - \!\vec{A}\vX}_2^2 \biggiv \vec{y}; \psi^i\} \nonumber \\
&= \vec{y}\tran\vec{y} - \vec{y}\tran\!\vec{A}\hvec{x} + \hat{\E}\{\vX\tran\!\vec{A}\!\tran\vec{A}\vX \biggiv \vec{y}; \psi^i\} \\
&= \vec{y}\tran\vec{y} - \vec{y}\tran\!\vec{A}\hvec{x} + \tr(\vec{A}\tran\!\vec{A}\vec{\Sigma}) + \hvec{x}\tran\!\vec{A}\tran\!\vec{A}\hvec{x} \\
\label{eq:EMAWGN3}
&= \norm{\vec{y}- \vec{A}\hvec{x}}_2^2 + \tr(\vec{A}\tran\!\vec{A}\vec{\Sigma}).
\end{align}
Here, $\vec{\Sigma}$ is the posterior covariance matrix of $\vX$, which---based on our assumptions---is diagonal with $[\vec{\Sigma}]_{nn}=\mu^x_n$.
Plugging in \eqref{EMAWGN3} into \eqref{EMAWGN2}, we obtain the EM update \eqref{EMAWGNfinal}.
}

\section{Derivation of Laplacian likelihood quantities} \label{app:Lap}
\subsection{Laplacian likelihood steps for sum-product GAMP} \label{app:MMSE_Lap}

Inserting the Laplacian likelihood \eqref{LAP} into the GAMP-approximated posterior \eqref{gamppostz}, the posterior mean in line (R5) of \tabref{gamp} becomes (removing the $m$ subscript for brevity)
\begin{equation}
\label{eq:Lapzhat1}
\hat{z} 
\defn \E\{\Z \giv \P \!=\! \hat{p};\mu^p\} 
\!=\!\frac{1}{C}\int_{z} z\,\mc{L}(z;y;\psi)\mc{N}(z;\hat{p},\mu^p)
\end{equation}
where the scaling constant $C$ is calculated as
\begin{eqnarray}
C &=& \int_{z} \mc{L}(z; y, \psi) \mc{N}(z; \hat{p}, \mu^p)\\
&=& \int_{z'} \mc{L}(z'; 0, \psi) \mc{N}(z' ; \hat{p} - y, \mu^p)\label{eq:Cdefsstep}\\
&=& \underbrace{\frac{\psi}{2}\int_{-\infty}^{0} \mc{N}(z; \tilde{p}, \mu^p) e^{\psi z} dz}_{\displaystyle \defn \underline{C}} 
+  \underbrace{\frac{\psi}{2}\int_{0}^{\infty} \mc{N}(z; \tilde{p}, \mu^p) e^{-\psi z} dz}_{\displaystyle \defn \overline{C}} \nonumber\\[-5mm]&&\label{eq:Cdefs}
\end{eqnarray}
where $\tilde{p}\defn \hat{p}-y$.
The expressions for $\underline{C}$ and $\overline{C}$ reported in \eqref{CL}-\eqref{CU} result after completing the square inside the exponential terms in the integrands in \eqref{Cdefs} and simplifying.  

Following similar techniques (\ie shifting $z$ by $y$ and splitting the integral), it can be shown that \eqref{Lapzhat1} becomes
\begin{align}
\label{eq:Lapzhat2}
\hat{z} \!=\! y +\frac{\underline{C}}{C} \!\int_{z}\! z \mc{N}_-(z;\tilde{p},\mu^p) + \frac{\overline{C}}{C} \int_{z}\! z \mc{N}_+(z;\tilde{p},\mu^p) , 
\end{align}
where $\mc{N}_+(\cdot)$ is defined in \eqref{nng} and where $\mc{N}_-(x;a,b^2)$ is the pdf that results from taking a Gaussian with mean $a$ and variance $b^2$, truncating its support to $x \in (-\infty,0]$, and normalizing.  
Supposing that $\U \sim \mc{N}(a,b^2)$, \cite{Johnson:Book:95} shows that
\begin{align}
\label{eq:trun_gaus}
\E\{\U \giv \U > 0\} &= \int_u \textb{u} \mc{N}_+(u;a,b^2) = a + b h(-\tfrac{a}{b}),\\
\E\{\U \giv \U < 0\} &= \int_u u \mc{N}_-(u;a,b^2)= a - b h(\tfrac{a}{b}),
\label{eq:trun_gaus-}
\end{align}
where $h(\cdot)$ is defined in \eqref{h}.
Inserting \eqref{trun_gaus} and \eqref{trun_gaus-} into \eqref{Lapzhat2} yields the posterior mean expression in \eqref{MMSELapmean}.

To calculate the posterior variance $\mu^z$ used in line (R6) of \tabref{gamp}, we begin with
\begin{align}
\label{eq:Lapzvar1}
&\E\{\Z^2\giv\P \!=\! \hat{p};\mu^p\} = \frac{1}{C}\int_{z} z^2\mc{L}(z;y;\psi)\mc{N}(z;\hat{p},\mu^p) \\
&= \frac{1}{C}\int_{z'} (z'+y)^2\mc{L}(z';0;\psi)\mc{N}(z';\tilde{p},\mu^p)\\
&= 2y(\hat{z} - y) + y^2  + \frac{1}{C}\int_z z^2\mc{L}(z;0;\psi)\mc{N}(z;\tilde{p},\mu^p) \\
&= 2y\hat{z} - y^2 
+ \frac{\underline{C}}{C} \int_z z^2 \mc{N}_-(z;\tilde{p},\mu^p) + \frac{\overline{C}}{C} \int_z z^2 \mc{N}_+(z;\tilde{p},\mu^p).
\label{eq:Lapzvar2}
\end{align}
Given that $\U \sim \mc{N}(a,b^2)$, \cite{Johnson:Book:95} shows that
\begin{align}
\E\{\U^2 \giv \U \!>\! 0\} &= \var\{\U\giv\U \!>\! 0 \} + \E\{\U\giv\U \!>\!0\}^2 \nonumber\\
\label{eq:trun_gaus_var}
&= b^2g(-\tfrac{a}{b}) + \big(a + b h(-\tfrac{a}{b})\big)^2, \\
\E\{\U^2 \giv \U \!<\! 0\} &= \var\{\U\giv\U \!<\! 0 \} + \E\{\U\giv \U \!<\!0\}^2 \nonumber\\
\label{eq:trun_gaus-_var}
&= b^2g(\tfrac{a}{b}) + \big(a - b h(\tfrac{a}{b})\big)^2,
\end{align}
where $g(\cdot)$ is defined in \eqref{g}.
Inserting \eqref{trun_gaus_var} and \eqref{trun_gaus-_var} into \eqref{Lapzvar2} and noting that $\var\{\Z\giv\P \!=\! \hat{p};\mu^p\} = \E\{\Z^2\giv\P \!=\! \hat{p};\mu^p\} - \E\{\Z|\P \!=\! \hat{p};\mu^p\}^2$, we obtain \eqref{MMSELapvar}.

\subsection{EM update for Laplacian rate} 				\label{app:lap_upd}
Inserting the Laplacian likelihood \eqref{LAP} into \eqref{MLLaprate}, we see that the EM update for the Laplacian rate parameter $\psi$ becomes.
\begin{eqnarray}
\psi^{i+1} 
&=& \argmax_{\psi} \textb{\sum_{m=1}^M}\hat{\E}\{ \ln \mc{L}(y_m;\textb{\vec{a}\tran_m\vX},\psi) \biggiv \vec{y}; \psi^{i} \} \\
\label{eq:ML_Laprate1}
&=& \argmax_{\psi} M \ln \psi - \psi \sum_{m=1}^M \hat{\E}\{ |\textb{\vec{a}\tran_m\vX} - y_m| \biggiv \vec{y}; \psi^{i} \}. \qquad
\end{eqnarray}
Zeroing the derivative of the objective in \eqref{ML_Laprate1} w.r.t.\ $\psi$ yields the update \eqref{EM_Lap}.
The expectation in \eqref{ML_Laprate1} can be written as
\textb{
\begin{equation}
\hat{\E}\{ |\vec{a}\tran_m \vX - y_m | \biggiv \vec{y}; \psi^i\}
\label{eq:EMAWLN1}
= \!\int_{\vec{x}} |\vec{a}\tran_m\vec{x} \!-\! y_m|\, f_{\vX|\vY}(\vec{x}|\vec{y}; \psi^i),
\end{equation}
where $f_{\vX|\vY}(\vec{x}|\vec{y}; \psi^i)$ is taken to be the product of the approximated GAMP marginal posteriors in \eqref{gamppost}.
}

\textb{
In the large system limit, the central limit theorem implies that $\Z_m \!\defn\! \vec{a}\tran_m\vX$, when conditioned on $\vY\!=\!\vec{y}$, is 
$\mc{N}(\vec{a}\tran_m \hvec{x}, \sum_{n=1}^N \vec{a}^2_{mn}\mu^x_n)$,
yielding the approximation 
\begin{align}
&\hat{\E}\{ |\vec{a}\tran_m \vX - y_m | \biggiv \vec{y}; \psi^i\}  \nonumber \\
&\approx \int_{z_m} |z_m - y_m| \,\mc{N}(z_m; \vec{a}\tran_m \hvec{x}, \sum_{n=1}^N \vec{a}^2_{mn}\mu^x_n) \\
&= \int_{z_m^\prime} |z_m^\prime| \,\mc{N}(z_m'; \vec{a}\tran_m\hvec{x}- y_m, \sum_{n=1}^N \vec{a}^2_{mn}\mu^x_n).
\end{align}
Defining $\tilde{z}_m \defn \vec{a}\tran_m\hvec{x} - y_m$, and using a derivation similar to that used for \eqref{Lapzhat2}, leads to \eqref{Lapquant}.
}


\section{Derivation of NNGM-GAMP quantities} 				\label{app:NNGM}
\subsection{BNNGM prior steps for sum-product GAMP}		        \label{app:MMSE_NNGM}
Inserting the Bernoulli NNGM prior \eqref{mmsepri} into the GAMP approximated posterior \eqref{gamppost}, the posterior mean in line (R13) of \tabref{gamp} becomes (removing the $n$ subscript for brevity)
\begin{align}
\label{eq:mom_def}
&\hat{x} \defn \E\{\X\giv\R = \hat{r}; \mu^r\} 
= \int_x x \, f_{\X|\R}(x| \hat{r}; \mu^r) \\
&= \frac{1}{\zeta} \int_+ \!x\,\mc{N}(x;\hat{r},\mu^r)\Big( (1-\tau) \delta(x) \!+\! \tau \sum_{\ell=1}^L \omega_\ell \mc{N}_+(x;\theta_\ell,\phi_\ell)\Big) \nonumber\\
\label{eq:mom_calc}
&= \frac{\tau}{\zeta}\sum_{\ell=1}^L \omega_\ell\int_+ x\mc{N}(x;\hat{r},\mu^r) \mc{N}_+(x;\theta_\ell,\phi_\ell),
\end{align}
where $\zeta \defn \int_x f_\X(x)\mc{N}(x;\hat{r},\mu^r)$ is a scaling factor.  
Using the Gaussian-pdf multiplication rule,\footnote{
$\mc{N}(x;a,A)\mc{N}(x;b,B) \!=\! \mc{N}\left(x; \frac{a/A + b/B}{1/A + 1/B},\frac{1}{1/A + 1/B}\right)\mc{N}(a;b,A\!+\!B).$}
we get
\begin{equation}
 \label{eq:mom_calc2}
\hat{x} = \frac{\tau}{\zeta}\sum_{\ell=1}^L \frac{\omega_\ell\mc{N}(\hat{r};\theta_\ell,\mu^r + \phi_\ell)}{\Phi_c(-\theta_\ell/\sqrt{\phi_\ell})} \int_+ x\, \mc{N}(x;\gamma_\ell,\nu_\ell),
\end{equation}
with $\gamma_\ell$ and $\nu_\ell$ defined in \eqref{gamma} and \eqref{nu}, respectively. 

Using similar techniques, the scaling factor
\begin{equation}
\label{eq:norm_begin}
\zeta = \int_+ \mc{N}(x;\hat{r},\mu^r)\Big( (1-\tau) \delta(x) + \tau \sum_{\ell=1}^L \omega_\ell \mc{N}_+(x;\theta_\ell,\phi_\ell)\Big)
\end{equation}
can be shown to be equivalent to \eqref{norm}.
Finally, using the mean of a truncated Gaussian \eqref{trun_gaus}, and inserting \eqref{norm} into \eqref{mom_calc2},  we get the NNGM-GAMP estimate \eqref{mean_final}.

To calculate the variance of the GAMP approximated posterior \eqref{gamppost}, we note that
\begin{align}
\mu^x &\defn \var\{\X\giv\R=\hat{r};\mu^r\} \nonumber\\
\label{eq:gamp_post_var}
&= \int_+ x^2 f_{x|\R}(\X|\hat{r}; \mu^r) - \E\{\X\giv\R =\hat{r};\mu^r\}^2.
\end{align}
Following \eqref{mom_def}-\eqref{mom_calc2} and using the Gaussian-pdf multiplication rule, we find the second moment to be
\begin{equation}
\label{eq:sec_mom}
\int_+ x^2 f_{\X|\R}(x|\hat{r};\mu^r) 
 \!=\! \frac{\tau}{\zeta} \sum_{\ell =1}^L \frac{\beta_\ell}{\Phi_c(\alpha_\ell)} \int_+ x^2 \mc{N}\left(x; \gamma_\ell, \nu_\ell \right),
\end{equation}
where $\beta_\ell$ and $\alpha_\ell$ are given in \eqref{beta} and \eqref{alpha}, respectively.

Leveraging the second moment of a truncated Gaussian \eqref{trun_gaus_var} in \eqref{sec_mom}, and then inserting \eqref{mean_final} and \eqref{sec_mom} into \eqref{gamp_post_var}, we obtain the NNGM-GAMP variance estimate \eqref{var_final}.  

\subsection{EM updates of NNGM parameters} 			\label{app:NNGM_upd}

We first derive the EM update for $\theta_k$, the $k^{th}$ component location, given the previous parameter estimate $\vec{q}^i$.
The maximizing value of $\theta_k$ in \eqref{MLtheta}  is necessarily a value of $\theta_k$ that zeros the derivative of the sum, \ie that satisfies\footnote{By employing the Dirac approximation $\delta(x)=\mc{N}(x;0,\varepsilon)$ for fixed arbitrarily small $\varepsilon>0$, the integrand and its derivative w.r.t $\theta_k$ become continuous, justifying the exchange of differentiation and integration via the Leibniz integration rule.  
We apply the same reasoning for all exchanges of differentiation and integration in the sequel.}
\begin{align}
\frac{d}{d\theta_k}\sum_{n=1}^N \int_{x_n} \!f_{\X|\R}\big(x_n|\hat{r}_n;\mu^r_n, \vec{q}^i\big) \ln f_\X\left (x_n;\theta_k, \vec{q}^i\remove{\theta_k}\right) \!&=\! 0\\
\label{eq:EMtheta}
\Leftrightarrow \sum_{n=1}^N \int_{x_n} \!f_{\X|\R}\big(x_n|\hat{r}_n; \mu^r_n, \vec{q}^i\big) \frac{d}{d\theta_k} \ln f_\X\left (x_n;\theta_k, \vec{q}^i\remove{\theta_k}\right) \!&=\! 0.
\end{align}

For all $x_n \geq 0$, the derivative in \eqref{EMtheta} can be written as 
\begin{equation}
\label{eq:thetakdertrue}
\frac{d}{d\theta_k}\ln f_\X(x_n; \theta_k, \vec{q}^i_{\setminus \theta_k}) 
= \frac{\tfrac{d}{d \theta_k} \tau^i \omega_k^i \tfrac{\mc{N}(x_n;\theta_k;\phi_k^i)}{\Phi_c(-\theta_k/\sqrt{\phi_k^i})}}{f_\X(x_n; \theta_k, \vec{q}^i_{\setminus \theta_k})}.
\end{equation}
Because plugging \eqref{thetakdertrue} into \eqref{EMtheta} yields an intractable integral, we use the approximation\footnote{This approximation becomes more accurate as $\tfrac{d}{d\theta_k}\Phi_c \big(-\theta_k/{\sqrt{\phi_k}}\big)$ tends to zero, \ie when $\theta_k/\sqrt{\phi_k}$ gets large\textb{,
which was observed for the real-world experiments considered in \secref{results}.}}
$\Phi_c(-\theta_k/\sqrt{\phi_k^i}) \approx \Phi_c(-\theta^i_k/\sqrt{\phi_k^i})$, yielding
\vspace{-3mm}

\small
\begin{align}
&\frac{d}{d\theta_k}\ln f_\X(x_n; \theta_k, \vec{q}^i_{\setminus \theta_k}) 
\!=\! \left(\frac{x_n - \theta_k}{\phi_k^i}\right) \label{eq:thetakder} \\
&\times \!\!
\frac{\tau^i\omega_k^i \mc{N}(x_n;\theta_k,\phi_k^i)/\Phi_c\big(-\theta^i_k/\sqrt{\phi_k^i}\big)}
{(1\!-\!\tau^i)\delta(x_n) \!\!+\! \tau^i(\omega_k^i \mc{N}_+ (x_n; \theta_k,\phi_k^i) \!\!+\!\! \sum_{\ell \neq k} \!\omega_\ell^i \mc{N}_+(x_n;\theta_\ell^i,\phi_\ell^i))}.
	\nonumber
\end{align}
\normalsize
\vspace{-2mm}

\noindent
We also note that \eqref{thetakder} is zero at $x_n =0$ due to the Dirac delta function in the denominator.  

Now, plugging in \eqref{thetakder} and the approximated GAMP posterior $f_{\X|\R}(x_n|\hat{r}_n;\mu^r_n, \vec{q}^i)$ from \eqref{gamppost},
integrating \eqref{EMtheta} separately over $[\epsilon,\infty)$ and its complement, and taking $\epsilon\rightarrow 0$, we find that the $(-\infty,\epsilon)$ portion vanishes, giving the necessary condition
\begin{equation}
\label{eq:thetak2}
\sum_{n=1}^{N} \int_+  \!\!
\frac{\hat{p}(x_n|x_n \neq 0,\vec{y};\vec{q}^i)\omega_k^i\tfrac{\mc{N}(x_n;\theta_k,\phi_k^i)}{\Phi_c(-\theta^i_k/\sqrt{\phi_k^i})}(x_n - \theta_k)}
{\zeta_n\big(\omega_k^i \mc{N}_+ (x_n; \theta_k,\phi_k^i) + \sum_{\ell \neq k} \omega_\ell^i \mc{N}_+(x_n;\theta_\ell^i,\phi_\ell^i)\big)}  \!=\! 0,
\end{equation}
where $\hat{p}(x_n|x_n \!\neq\! 0,\vec{y};\vec{q}^i) \defn f_{\X|\R}(x_n|\hat{r}_n,x_n \!\neq\! 0;\mu^r_n,\vec{q}^i)$.
Since this integral cannot be evaluated in closed form, we apply the approximation $\mc{N}(x_n;\theta_k,\phi_k^i) \approx \mc{N}(x_n;\theta_k^i,\phi_k^i)$ in both the numerator and denominator, and subsequently exploit the fact that, for $x_n \geq 0$, 
$\hat{p}(x_n|x_n \neq0,\vec{y};\vec{q}^i)
=\mc{N} (x_n; \hat{r}_n, \mu^r_n) 
\sum_{\ell} \omega_\ell^i \mc{N}_+(x_n;\theta_\ell^i,\phi_\ell^i)$ 
from \eqref{gamppost} to cancel terms, where we obtain the necessary condition
\begin{equation}
\label{eq:thetak3}
\sum_{n=1}^N \int_+ \frac{\omega_k^i\mc{N} (x_n; \hat{r}_n, \mu^r_n)
\mc{N}_+(x_n;\theta_k^i,\phi_k^i)}{\zeta_n}
(x_n - \theta_k) = 0.
\end{equation}
Now using the Gaussian-pdf multiplication rule, we get 
\begin{equation}
\label{eq:thetak4}
\sum_{n=1}^N \frac{\beta_{n,k}}{\Phi_c(\alpha_{n,k})} \int_+ \mc{N}(x_n;\gamma_{n,k},\nu_{n,k})
(x_n - \theta_k) = 0.
\end{equation}
Following similar techniques as in \appref{MMSE_NNGM} and noting that $\beta_{n,k} = \pi_n \overline{\beta}_{n,k}$, we see that the update $\theta_k^{i+1}$ in \eqref{EMthetafinal} is the value of $\theta_k$ that satisfies \eqref{thetak4}.

Similarly, the maximizing value of $\phi_k$ in \eqref{MLphi} is necessarily a value of $\phi_k$ that zeroes the derivative, i.e.,
\begin{equation}
\label{eq:phik}
\sum_{n=1}^{N} \int_{x_n} f_{\X|\R}(x_n|\hat{r}_n;\mu^r_n,\vec{q}^i) \frac{d}{d \phi_k}\ln f_\X(x_n;\phi_k,\vec{q}_{\setminus \phi_k}^i) = 0.
\end{equation}
Using the prior given in \eqref{mmsepri}, and simultaneously applying the approximation $\Phi_c(-\theta_k^i/\sqrt{\phi_k}) \approx \Phi_c(-\theta^i_k/\sqrt{\phi_k^i})$, we see that the derivative in \eqref{phik} can be written as
\vspace{-3mm}

\small
\begin{align}
&\frac{d}{d\phi_k}\ln f_\X(x_n; \phi_k, \vec{q}^i_{\setminus \phi_k}) 
\!=\!\frac{1}{2} \left(\frac{(x_n - \theta_k^i)^2}{\phi_k^2}-\frac{1}{\phi_k}\right)  \label{eq:phikder} \\
&\times \!\!
\frac{\tau^i\omega_k^i \mc{N}(x_n;\theta_k,\phi_k^i)/\Phi_c(-\theta^i_k/\sqrt{\phi_k^i})}
{(1\!-\!\tau^i)\delta(x_n) \!+\! \tau^i(\omega_k^i \mc{N}_+ (x_n; \theta^i_k,\phi_k) \!\!+\!\! \sum_{\ell \neq k} \!\omega_\ell^i \mc{N}_+(x_n;\theta_\ell^i,\phi_\ell^i))} .
	\nonumber
\end{align}
\normalsize
\vspace{-2mm}

\noindent
Integrating \eqref{phik} separately over $(-\infty,\epsilon)$ and $[\epsilon,\infty)$, and taking $\epsilon\rightarrow 0$, we see that the $(-\infty,\epsilon)$ portion vanishes, giving
\begin{align}
&\sum_{n=1}^{N}\int_+ \!\!
\frac{\hat{p}(x_n|x_n \!\neq\! 0,\vec{y};\vec{q}^i)\omega_k^i\mc{N}(x_n;\theta_k^i,\phi_k)/\Phi_c(-\theta^i_k/\!\sqrt{\phi_k^i})}
{\zeta_n(\omega_k^i \mc{N} (x_n; \theta_k^i,\phi_k) + \sum_{\ell \neq k} \omega_\ell^i \mc{N}(x_n;\theta_\ell^i,\phi_\ell^i))} \nonumber \\
\label{eq:phik2}
&\times \bigg(\frac{(x_n - \theta_k^i)^2}{\phi_k}-1\bigg) = 0. 
\end{align}
Again, this integral is difficult to compute, so we apply the approximation $\mc{N}(x_n;\theta_k,\phi_k^i) \approx \mc{N}(x_n;\theta_k^i,\phi_k^i)$ in both the numerator and denominator. 
After some cancellation (as in \eqref{thetak2}), we get the necessary condition
\vspace{-3mm}

\small
\begin{align}
\label{eq:phik3}
\sum_{n=1}^N \!\int_+\! \frac{\mc{N} (x_n; \hat{r}_n, \mu^r_n)
\omega_k^i \mc{N}_+(x_n;\theta_k^i,\phi_k^i)}{\zeta_n}
\!\left(\frac{(x_n-\theta_k^i)^2}{\phi_k} -1\!\right) \!=\! 0.
\end{align}
\normalsize
To find the value of $\phi_k$ that solves \eqref{phik3}, we expand $(x_n - \theta_k^i)^2 = x_n^2 - 2 x_n\theta_k^i + (\theta_k^i)^2$ and apply the Gaussian-pdf multiplication rule, yielding
\begin{equation}
\label{eq:phik4}
\sum_{n=1}^N \frac{\beta_{n,k}}{{\Phi_c(\alpha_{n,k})}} \!\int_+ \!\mc{N}(x_n;\gamma_{n,k},\nu_{n,k})
\!\left(\!\frac{x_n^2 \!-\! 2 x_n\theta_k^i \!+\! (\theta_k^i)^2}{\phi_k} -\! 1 \!\right) \!\!=\! 0.
\end{equation}
\normalsize
\vspace{-2mm}

\noindent
Using similar techniques as in \appref{MMSE_NNGM} and simplifying, we see that $\phi_k^{i+1}$ in \eqref{EMSphifinal} is the value of $\phi_k$ that solves \eqref{phik4}.  

Finally, we calculate the EM update in \eqref{MLomega} for positive $\vec{\omega}$ under the pmf constraint $\sum_{k=1}^L \omega_k = 1$
by solving the unconstrained optimization problem 
$\max_{\vec{\omega},\xi} J(\vec{\omega},\xi)$, where $\xi$ is a Lagrange multiplier and 
\begin{align}
&J(\vec{\omega},\xi) \defn\!
\sum_{n=1}^N \hat{\E}\big\{\!\ln f_\X(\X_n; \vec{\omega}, \vec{q}^i_{\setminus \vec{\omega}}) \biggiv \vec{y}; \vec{q}^i\big\}
\!-\! \xi\bigg(\sum_{\ell=1}^L \omega_\ell \!-\! 1 \bigg).
\end{align}
First, we set $\frac{d}{d\omega_k} J(\vec{\omega},\xi)=0$,
which yields 
\begin{align}
\label{eq:omegak1}
	\sum_{n=1}^N \int_{x_n} \!\!\frac{f_\X(x_n;\vec{q}^i) \mc{N}(x_n;\hat{r}_n,\mu^r_n)}
			{\zeta_n}
		\frac{d}{d\omega_k} \ln f_\X(x_n; \vec{\omega}, \vec{q}^i_{\setminus \vec{\omega}}) &= \xi \!
\end{align}
where, for non-negative $x_n$,
\begin{align}
\label{eq:omegader}
&\frac{d}{d\omega_k} \ln f_\X\big(x_n; \vec{\omega}, \vec{q}^i\remove{\vec{\omega}}\big) = 
\mbox{}\frac{\tau^i \mc{N}_+ (x_n;\theta_k^i,\phi_k^i)}{ f_\X\big(x_n; \vec{\omega}, \vec{q}^i\remove{\vec{\omega}}\big)}.
\end{align}
Inserting \eqref{omegader} into \eqref{omegak1}, we get
\begin{eqnarray}
\sum_{n=1}^N \int_+\!\!\frac{f_\X(x_n;\vec{q}^i) \mc{N}(x_n;\hat{r}_n,\mu^r_n)} {\zeta_n}
		\frac{\tau^i \mc{N}_+(x_n;\theta_k^i,\phi_k^i)}
			{f_\X(x_n;\vec{\omega},\vec{q}_{\setminus\vec{\omega}}^i)} &=& \xi . ~~\qquad
\end{eqnarray}
As in \eqref{thetak2} and \eqref{phik2}, the above integral is difficult to evaluate,
and so we apply the additional approximation $\vec{\omega}\approx \vec{\omega}^i$, which reduces the previous equation to 
\begin{equation}
\xi 
= \sum_{n=1}^N \int_+ \frac{ \tau^i \mc{N}_+(x_n;\theta_k^i,\phi_k^i)
				\mc{N}(x_n;\hat{r}_n,\mu^r_n) } {\zeta_n} .
				\label{eq:xi1}
\end{equation}
We then multiply both sides by $\omega_k^i$ for $k=1,\dots,L$,  and sum over $k$.
Leveraging the fact $1=\sum_k \omega_k^i$, and simplifying, we obtain the equivalent condition 
\begin{eqnarray}
\xi 
&=& \sum_{n=1}^N \int_+ \frac{ \tau^i \sum_{k=1}^L \omega_k^i \mc{N}_+(x_n;\theta_k^i,\phi_k^i) \mc{N}(x_n;\hat{r}_n,\mu^r_n) } {\zeta_n} \qquad \\
&=& \sum_{n=1}^N \frac{\tau^i}{\zeta_n} \sum_{k=1}^L \beta_{n,k} \int_+ \frac{\mc{N}(x_n;\gamma_{n,k},\phi_{n,k})}{{\Phi_c(\alpha_{n,k})}} = \sum_{n=1}^N \pi_n .  \label{eq:xi2}
\end{eqnarray}
Plugging \eqref{xi2} into \eqref{xi1} and multiplying both sides by $\omega_k$,
the derivative-zeroing value of $\omega_k$ is seen to be
\begin{equation}
\omega_k
\!=\! \frac{ \sum_{n=1}^N \!\int_+ \!\! \tau^i \omega_k \mc{N}_+(x_n;\theta_k^i,\phi_k^i)\mc{N}(x_n;\hat{r}_n,\mu^r_n) / \zeta_n }
	{\sum_{n=1}^N \pi_n } ,
				\label{eq:omegak2}
\end{equation}
where, if we use $\omega_k\approx \omega_k^i$ on the right of \eqref{omegak1}, then we obtain the approximate EM update $\omega_k^{i+1}$ in \eqref{EMSomegafinal}.

\bibliographystyle{ieeetr}
\bibliography{macros_abbrev,books,misc,sparse,machine,hsi}
\end{document}